\documentclass[aps,prc,superscriptaddress,twoside,twocolumn,nofootinbib,10pt,
showpacs,floatfix]{revtex4-1}

\usepackage{amsmath,amssymb}
\usepackage{graphicx,bm}
\usepackage{epstopdf}
\usepackage{ulem}
\usepackage[usenames]{color}
\usepackage{float}
\usepackage{color}

\def\slash#1{\not\!#1}

\allowdisplaybreaks

\begin{document}

\author{Yusuke Takeda}
\affiliation{Department of Physics, Nagoya University, Nagoya, 464-8602, Japan}

\author{Youngman Kim}
\affiliation{Rare Isotope Science Project, Institute for Basic Science, Daejeon 305-811, Korea}

\author{Masayasu Harada}
\affiliation{Department of Physics, Nagoya University, Nagoya, 464-8602, Japan}

\title{Catalysis of partial chiral symmetry restoration by  Delta matter}

\begin{abstract}
We study the phase structure of dense hadronic matter including
 $\Delta(1232)$ as well as $N(939)$
based on the parity partner structure, where  the baryons have their chiral partners with a certain amount of chiral invariant masses.
We show that, in symmetric matter, $\Delta$ enters into matter in the density region of about one to four times of normal nuclear matter density, $\rho_B \sim 1 $ - $4 \rho_0$.
The onset density of $\Delta$ matter depends on the chiral invariant mass of $\Delta$, $m_{\Delta 0}$: The lager $m_{\Delta 0}$, the bigger the onset density.
 The $\Delta$ matter of $\rho_B \sim 1$ - $4 \rho_0$
 is unstable due to the existence of $\Delta$, and the stable $\Delta$-nucleon matter is realized at about $\rho_B \sim 4\,\rho_0$, i.e., the  phase transition from nuclear matter to $\Delta$-nucleon matter is of first order for small $m_{\Delta 0}$, and it is of second order for large $m_{\Delta 0}$.
We find that, associated with the phase transition, the chiral condensate changes very rapidly, i.e., the chiral symmetry restoration is accelerated by $\Delta$ matter.
As a result of the accelerations, there appear $N^\ast(1535)$ and $\Delta(1700)$, which are the chiral partners to $N(939)$ and $\Delta(1232)$, in high density matter, signaling the partial chiral symmetry restoration.
Furthermore, we find that complete chiral symmetry restoration itself is delayed by $\Delta$ matter.
We also calculate the effective masses, pressure and symmetry energy to study how the transition to $\Delta$ matter affects such physical quantities.
We observe that the physical quantities change drastically at the transition density.

\end{abstract}


\maketitle

\section{Introduction}
Studying the origin of hadron masses is one of interesting problems in hadron physics.
As it is well known,  the mass of current quarks could explain roughly 2$\%$ of  the nucleon mass.
A standard folklore in nuclear physics states that
the nucleon mass other than that from the current quarks could be explained in terms of spontaneous chiral symmetry breaking.
However, in the parity doublet model it becomes even more interesting as the nucleon mass gets an additional contribution from  a so-called chiral invariant mass.
This structure can be formulated in models with parity doublet structure \cite{Detar:1988kn,Jido:2001nt}.
It is interesting to investigate the chiral invariant mass portion of the nucleon mass compared to that from chiral symmetry breaking.
Since the chiral symmetry is expected to be partially restored in dense matter,  a part of hadron  masses originated from the spontaneous chiral symmetry breaking is changed. This change will provide a clue for elucidating the origin of hadron masses.

Dense nuclear matter, which is intimately related to heavy ion collisions, nuclear structure, and neutron stars, has served as a testing ground for our understanding
of non-perturbative QCD in hadron and nuclear physics.
Dense matter with large neutron-proton number asymmetry is important to understand neutron star properties. Isospin asymmetric dense matter has been even more highlighted thanks to existing and forthcoming rare isotope accelerator facilities.

There are many extensive studies of  dense matter in the parity doublet models~\cite{Hatsuda:1988mv, Zschiesche:2006zj, Dexheimer:2007tn, Gallas:2009qp, Sasaki:2010bp, Gallas:2011qp, Steinheimer:2011ea, Benic:2015pia, Motohiro:2015}.
It was shown in Ref.~\cite{Motohiro:2015} that the phase structure depends on the value of the chiral invariant mass.
The density dependence of the effective nucleon mass was also presented, and it was pointed that the masses change, reflecting the partial chiral symmetry restoration in nuclear matter.
It will be also interesting to ask “What happens to the masses of other hadrons in matter associated with the partial restoration of chiral symmetry.”

In the present study, we focus on the role of the $\Delta$ baryon in terrestrial dense matter that could be created in heavy ion collisions at a few hundreds MeV~\cite{Li:1997yh,Lavagno:2012bn}.
In Ref.~\cite{Li:1997yh}, it was shown that, in symmetric matter, the $\Delta$ matter appears when the coupling of $\sigma$ mean field to $\Delta$ is larger than that to nucleon, and that the on-set density strongly depends on the saturation properties.
In Ref.~\cite{Lavagno:2012bn},
the $\Delta$ matter in asymmetric matter was studied with focusing on the stability of the system. It was shown that there can be instabilities when $\Delta$ matter appears.
There are also many works on the $\Delta$ baryons in neutron stars \cite{Xiang:2003qz,Schurhoff:2010ph,Drago:2014oja,Cai:2015hya,Zhu:2016mtc}.
It was pointed that the $\Delta$ baryon enters into the matter around two to three times of the normal nuclear matter density $\rho_0$.

In this work we study the chiral phase structure of dense matter that could be created during
heavy ion collisions at a few hundreds A MeV.
First we construct a parity doublet model
extended from the model in Ref.~\cite{Motohiro:2015} to include the $\Delta$ baryon following Ref.~\cite{Jido:1999hd,Jido:2001nt}.
In the present model, $N(1535)$ and $\Delta(1700)$ are included as chiral partners to $N(939)$ and $\Delta(1232)$, respectively, having certain amounts of chiral invariant masses.
We first investigate the dependence of matter constituents on the chiral invariant mass of $\Delta$, $m_{\Delta0}$, and show that the onset density of the appearance of $\Delta$ matter strongly depends on $m_{\Delta 0}$.  Then, we study the chiral phase structure and show that the partial chiral symmetry restoration is accelerated by $\Delta$ matter, while the chiral symmetry restoration itself is delayed.
In addition, we also calculate the effective masses, pressure and symmetry energy to study how the transition to $\Delta$ matter affects such physical quantities.

This paper is organized as follows:

In Sec. \ref{ePDM} we introduce the parity doublet model with the $\Delta$ baryon, and in Sec. \ref{effM}
we evaluate the effective $\Delta$ and nucleon masses in cold dense matter.
In Sec. \ref{deltaM}, we determine the model parameters and present our results focusing on the possibility of having the $\Delta$ baryons in dense matter
, i.e., a phase transition from nuclear matter to $\Delta$ matter.
Then, in Sec.~\ref{sec:chiral}, we investigate how the chiral structure is affected by $\Delta$ matter.
Finally, we summarize the present study with a brief discussion on the issues related to $\Delta$ matter
in Sec. \ref{summary}.

\section{Lagrangian for $\Delta$ Baryon }\label{ePDM}

In the present analysis,
nucleons are included using a hadronic model in Ref.~\cite{Motohiro:2015}.
We further include $\Delta$ baryon and its chiral partner based on the parity doublet structure following Ref.~\cite{Jido:1999hd,Jido:2001nt}.

For expressing $\Delta$ baryon and its chiral partner, we introduce two fields of spin $3/2$, $\psi_{1,2}^\mu$,
 to which we impose the constrains $\gamma_\mu\psi^\mu=0$ and $\partial_\mu\psi^\mu=0$, to reduce extra degrees of freedom.
To define the chiral transformation of the fields, we introduce
\begin{align}
\psi_{1r,2r}^\mu = \frac{1 + \gamma_5}{2} \psi_{1,2}^\mu \ ,  \quad
\psi_{1l,2l}^\mu = \frac{1 - \gamma_5}{2} \psi_{1,2}^\mu \ .
\end{align}
Their representations under the chiral $\mbox{SU}(2)_R \times \mbox{SU}(2)_L$ are given as
\begin{align}
(\psi_{1r,2l}^\mu)_{\alpha\beta}^\gamma  \in \left( \frac{1}{2}, 1 \right)  \ , \quad
(\psi_{1l,2r}^\mu )_\alpha^{\gamma\delta} \in \left( 1, \frac{1}{2} \right)  \ ,
\end{align}
where $\alpha$, $\beta$ are indices for $\mbox{SU}(2)_L$ and $\gamma$, $\delta$ for $\mbox{SU}(2)_R$.

We introduce the iso-singlet scalar meson $\sigma$ and the iso-triplet pion $\pi^A$ ($A=1,2,3$)  through the matrix field $M$ as
\begin{equation}
M = \sigma + i \sum_{A=1}^{3} \pi^A \tau^A \ ,
\end{equation}
where $\tau^A$ are the Pauli matrices.
The chiral representation of this $M$ is
\begin{equation}
M \in
\left(\frac{1}{2},\frac{1}{2} \right) \ .
\end{equation}

The Lagrangian for $\Delta$ baryon and its chiral partner is expressed as
		\begin{align}
			{\cal L}_{\Delta}&=(\bar\psi^\mu_{1r})_{\alpha\beta}^\gamma\left\{\sigma_{\mu\nu},i\slash D^{\gamma\gamma'}_{\beta\beta',\alpha\alpha'}\right\}
										(\psi^\nu_{1r})^{\gamma'}_{\beta'\alpha'}& \notag\\
							&+(\bar\psi^\mu_{1l})_{\alpha}^{\gamma\delta}\left\{\sigma_{\mu\nu},i\slash D^{\delta\delta',\gamma\gamma'}_{\alpha\alpha'}\right\}
										(\psi^\nu_{1l})^{\delta'\gamma'}_{\alpha'}& \notag\\
							&+(\bar\psi^\mu_{2r})_{\alpha}^{\gamma\delta}\left\{\sigma_{\mu\nu},i\slash D^{\delta\delta',\gamma\gamma'}_{\alpha\alpha'}\right\}
										(\psi^\nu_{2r})^{\delta'\gamma'}_{\alpha'}& \notag\\
							&+(\bar\psi^\mu_{2l})_{\alpha\beta}^\gamma\left\{\sigma_{\mu\nu},i\slash D^{\gamma\gamma'}_{\beta\beta',\alpha\alpha'}\right\}
										(\psi^\nu_{2l})^{\gamma'}_{\beta'\alpha'}& \notag\\
						&+m_0\Big((\bar\psi^\mu_{1r})^\gamma_{\alpha\beta}\sigma_{\mu\nu}(\psi^\nu_{2l})^\gamma_{\beta\alpha}
						-(\bar\psi^\mu_{1l})_\alpha^{\gamma\delta}\sigma_{\mu\nu}(\psi^\nu_{2r})_\alpha^{\delta\gamma}& \notag\\
						&+(\bar\psi^\mu_{2l})^\gamma_{\alpha\beta}\sigma_{\mu\nu}(\psi^\nu_{1r})^\gamma_{\beta\alpha}
						-(\bar\psi^\mu_{2r})_\alpha^{\gamma\delta}\sigma_{\mu\nu}(\psi^\nu_{1l})_\alpha^{\delta\gamma}\Big)& \notag\\
					&+a\left((\bar\psi^\mu_{1r})^\gamma_{\alpha\beta}M_\beta^\delta\sigma_{\mu\nu}(\psi^\nu_{1l})_\alpha^{\delta\gamma}
						+(\bar\psi^\mu_{1l})_\alpha^{\gamma\delta}(M^\dagger)_\beta^\delta\sigma_{\mu\nu}(\psi^\nu_{1r})_{\beta\alpha}^\gamma\right)& \notag\\
					&+b\left((\bar\psi^\mu_{2l})^\gamma_{\alpha\beta}M_\beta^\delta\sigma_{\mu\nu}(\psi^\nu_{2r})_\alpha^{\delta\gamma}
						+(\bar\psi^\mu_{2r})_\alpha^{\gamma\delta}(M^\dagger)_\beta^\delta\sigma_{\mu\nu}(\psi^\nu_{2r})_{\beta\alpha}^\gamma\right)\ ,&
		\end{align}
where
		\begin{align}
			(D_\mu)^{\gamma\gamma'}_{\beta\beta',\alpha\alpha'}&=\partial_\mu\delta^{\gamma\gamma'}\delta_{\beta\beta'}\delta_{\alpha'\alpha}
																	-i({\cal L}_\mu)^{\gamma\gamma'}\delta_{\beta\beta'}\delta_{\alpha'\alpha}&\notag \\
																	&-i\delta^{\gamma\gamma'}({\cal R}_\mu)_{\beta\beta'}\delta_{\alpha'\alpha}
																	-i\delta^{\gamma\gamma'}\delta_{\beta'\beta}({\cal R}_\mu)_{\alpha'\alpha}\ , &\\
			(D_\mu)^{\gamma\gamma',\delta\delta'}_{\alpha\alpha'}&=\partial_\mu\delta^{\gamma\gamma'}\delta^{\delta'\delta}\delta_{\alpha\alpha'}
																	-i({\cal L}_\mu)^{\gamma\gamma'}\delta^{\delta'\delta}\delta_{\alpha\alpha'}\notag \\
																	&-i\delta^{\gamma\gamma'}({\cal L}_\mu)^{\delta\delta'}\delta_{\alpha\alpha'}
																	-i\delta^{\gamma\gamma'}\delta^{\delta'\delta}({\cal R}_\mu)_{\alpha\alpha'}\, . &
		\end{align}
Here, the summation over repeated indices is understood.

We also introduce interactions to vector mesons in a similar way as for nucleons done in Ref.~\cite{Motohiro:2015} based on the hidden local symmetry~\cite{HLS}:
\begin{align}
{\cal L}_{{\rm vec}}=a_{\rho\Delta\Delta}\Big[
	&(\bar\psi^\mu_{1r})_{\alpha\beta}^\gamma\{\sigma_{\mu\nu},\gamma^\rho\}(\xi_R^\dagger\alpha_{\parallel\rho}\xi_R)_{\beta\beta'}(\psi^\nu_{1r})_{\beta'\alpha}^\gamma& \notag \\
	+&(\bar\psi^\mu_{1l})_{\alpha}^{\gamma\delta}\{\sigma_{\mu\nu},\gamma^\rho\}(\xi_L^\dagger\alpha_{\parallel\rho}\xi_L)^{\delta\delta'}(\psi^\nu_{1l})_{\alpha}^{\delta'\gamma}& \notag \\
	+&(\bar\psi^\mu_{1r})_{\alpha\beta}^\gamma\{\sigma_{\mu\nu},\gamma^\rho\}(\xi_R^\dagger\alpha_{\parallel\rho}\xi_R)_{\alpha'\alpha}(\psi^\nu_{1r})_{\beta\alpha'}^\gamma& \notag \\
	+&(\bar\psi^\mu_{1l})_{\alpha}^{\gamma\delta}\{\sigma_{\mu\nu},\gamma^\rho\}(\xi_L^\dagger\alpha_{\parallel\rho}\xi_L)^{\gamma'\gamma}(\psi^\nu_{1l})_{\alpha}^{\delta\gamma'}& \notag \\
	+&(\bar\psi^\mu_{1r})_{\alpha\beta}^\gamma\{\sigma_{\mu\nu},\gamma^\rho\}(\xi_L^\dagger\alpha_{\parallel\rho}\xi_L)^{\gamma\gamma'}(\psi^\nu_{1r})_{\beta\alpha}^{\gamma'}& \notag \\
	+&(\bar\psi^\mu_{1l})_{\alpha}^{\gamma\delta}\{\sigma_{\mu\nu},\gamma^\rho\}(\xi_R^\dagger\alpha_{\parallel\rho}\xi_R)_{\alpha\alpha'}(\psi^\nu_{1l})_{\alpha'}^{\delta\gamma}& \notag \\
	+&(\bar\psi^\mu_{2l})_{\alpha\beta}^\gamma\{\sigma_{\mu\nu},\gamma^\rho\}(\xi_R^\dagger\alpha_{\parallel\rho}\xi_R)_{\beta\beta'}(\psi^\nu_{2l})_{\beta'\alpha}^\gamma& \notag\\
	+&(\bar\psi^\mu_{2r})_{\alpha}^{\gamma\delta}\{\sigma_{\mu\nu},\gamma^\rho\}(\xi_L^\dagger\alpha_{\parallel\rho}\xi_L)^{\delta\delta'}(\psi^\nu_{2r})_{\alpha}^{\delta'\gamma}& \notag \\
	+&(\bar\psi^\mu_{2l})_{\alpha\beta}^\gamma\{\sigma_{\mu\nu},\gamma^\rho\}(\xi_R^\dagger\alpha_{\parallel\rho}\xi_R)_{\alpha'\alpha}(\psi^\nu_{2l})_{\beta\alpha'}^\gamma \notag \\
	+&(\bar\psi^\mu_{2r})_{\alpha}^{\gamma\delta}\{\sigma_{\mu\nu},\gamma^\rho\}(\xi_L^\dagger\alpha_{\parallel\rho}\xi_L)^{\gamma'\gamma}(\psi^\nu_{2r})_{\alpha}^{\delta\gamma'}& \notag \\
	+&(\bar\psi^\mu_{2l})_{\alpha\beta}^\gamma\{\sigma_{\mu\nu},\gamma^\rho\}(\xi_L^\dagger\alpha_{\parallel\rho}\xi_L)^{\gamma\gamma'}(\psi^\nu_{2l})_{\beta\alpha}^{\gamma'}& \notag \\
	+&(\bar\psi^\mu_{2r})_{\alpha}^{\gamma\delta}\{\sigma_{\mu\nu},\gamma^\rho\}(\xi_R^\dagger\alpha_{\parallel\rho}\xi_R)_{\alpha\alpha'}(\psi^\nu_{2r})_{\alpha'}^{\delta\gamma}\Big]& \notag\\
	+&a_{0\Delta\Delta}{\rm tr}[3\alpha_{\parallel\rho}]\Big[(\bar\psi^\mu_{1r})_{\alpha\beta}^\gamma\{\sigma_{\mu\nu},\gamma^\rho\}(\psi^\nu_{1r})_{\beta\alpha}^\gamma& \notag \\
	+&(\bar\psi^\mu_{1l})^{\gamma\delta}_\alpha\{\sigma_{\mu\nu},\gamma^\rho\}(\psi^\nu_{1l})^{\delta\gamma}_\alpha& \notag \\
	+&(\bar\psi^\mu_{2r})^{\gamma\delta}_\alpha\{\sigma_{\mu\nu},\gamma^\rho\}(\psi^\nu_{2r})^{\delta\gamma}_\alpha & \notag \\
	+&(\bar\psi^\mu_{2l})_{\alpha\beta}^\gamma\{\sigma_{\mu\nu},\gamma^\rho\}(\psi^\nu_{2l})_{\beta\alpha}^\gamma\Big]\, .&
\end{align}

\section{Effective masses}\label{effM}

We first construct a thermodynamic potential of  nuclear matter at mean field level including nucleons
of positive and negative parities in the model given in Ref.~\cite{Motohiro:2015},
with the baryon number chemical potential $\mu_B$ and the isospin chemical potential $\mu_I$.
Here, we assume that
$\mu_I$
is small and there is no pion condensation, and that the rotational invariance is not spontaneously broken.  Then, the relevant mean fields are $\bar{\sigma}$ for the scalar meson, $\bar{\omega}$ and $\bar{\rho}$ for the time components of the vector mesons, which are determined by minimizing the thermodynamic potential.

In nuclear matter, the existence of the mean fields of $\bar\omega$ and $\bar\rho$ changes the dispersion relation of the nucleon as
\begin{align}
E_n&=\sqrt{m_N^2+{\bf k}^2}+g_{\omega NN}\bar\omega -g_{\rho NN}\bar\rho & \notag\\
E_p&=\sqrt{m_N^2+{\bf k}^2}+g_{\omega NN}\bar\omega + g_{\rho NN}\bar\rho \ , &
\label{N energy}
\end{align}
where $E_{n,p}$ are the energies of a neutron and a proton measured from
their respective chemical potentials given by
\begin{align}
\mu_p & = \mu_B + \frac{1}{2} \mu_I \ , \notag\\
\mu_n & = \mu_B - \frac{1}{2} \mu_I \ .
\label{N chemical}
\end{align}
$g_{\omega NN}$ and $g_{\rho NN}$ are the couplings of $\omega$ and $\rho$ mesons to the nucleon.
Here, the nucleon mass $m_N$ is expressed by $\bar\sigma$ as
\begin{align}
m_N&=\frac{1}{2}\left(\sqrt{(g_1+g_2)^2\bar\sigma^2+4m^2_{N0}}+(g_1-g_2)\bar\sigma\right) \ ,
\label{N mass}
\end{align}
where $m_{N0}$ is the chiral invariant mass of the nucleon,
$g_1$ and $g_2$ are Yukawa couplings of the scalar meson to the nucleon.
The mass of its partner is given by
\begin{align}
m_{N^\ast}&=\frac{1}{2}\left(\sqrt{(g_1+g_2)^2\bar\sigma^2+4m^2_{N0}} - (g_1-g_2)\bar\sigma\right) \ ,
\label{N* mass}
\end{align}
which is the mass of $N^\ast(1535)$ as in Ref.~\cite{Motohiro:2015}.
Then, using the masses of $N(939)$ and $N^\ast(1535)$ as inputs, we determine the values of the $\sigma NN$ couplings for given value of the chiral invariant mass $m_{N0}$.
In Table~\ref{table:sigmaNN}, we list the values of $g_{1}$ and $g_{2}$ for several choices of $m_{N0}$.
\begin{table}[htbp]
\caption{Values of $\sigma N N$ coupling constants for several choices of the chiral invariant mass $m_{N0}$. }
\begin{tabular}{cccc}
\hline \hline
$m_{N 0}$ (MeV) &$500$ & $700$&$900$\\
\hline
$g_{1}$ & 9.03 & 7.82 & 5.97 \\
$g_{2}$ & 15.49 & 14.28 & 12.43 \\
\hline \hline
\label{table:sigmaNN}
\end{tabular}
\end{table}

In Fig. 1, we plot the density dependence of the mean field $\bar\sigma$ in symmetric nuclear matter with including only nucleons in matter.
\begin{figure}[htbp]
\vspace{-85pt}
\includegraphics[bb=0 0 480 360,width=10cm]{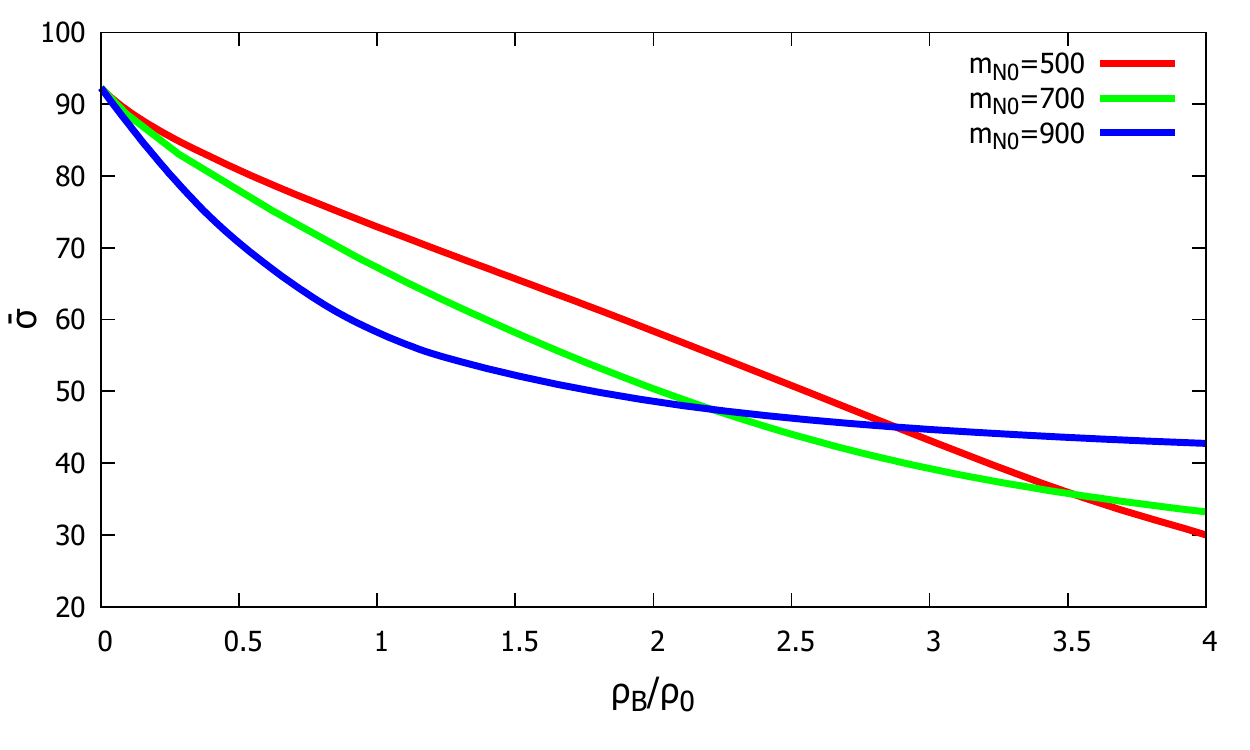}\\
\caption{Density dependence of $\bar\sigma$ in symmetric nuclear matter for $m_{N0}=500\,$MeV (red curve), $700$\,MeV (green curve) and $900\,$MeV (blue curve).
}
\end{figure}
This $\bar\sigma$ is interpreted as the pion decay constant in nuclear matter.
An experiment~\cite{Suzuki:2002ae} shows that the value of the pion decay constant at the normal nuclear matter density $\rho_0$ is about $0.8$ times of
that in vacuum~\cite{Kienle:2004hq}.
Figure 1 indicates that $m_{N0}=900$\,MeV leads to a too small value for $\bar\sigma$.
Then, in the following analysis, we use $m_{N0} = 500$-$700$\,MeV.

\begin{figure}[htbp]
\vspace{-85pt}
\includegraphics[bb=0 0 480 360,width=10cm]{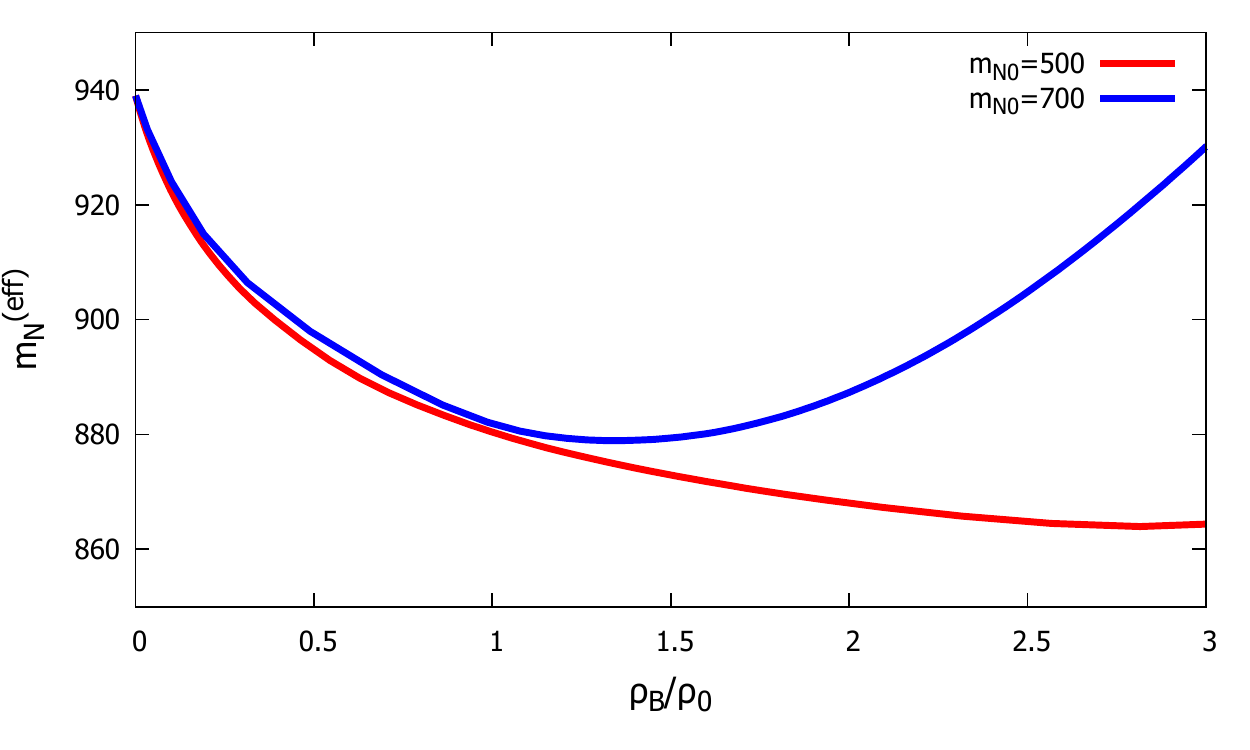}\\
\caption{Density dependence of effective mass of nucleon for $m_{N0}=500\,$MeV (red curve) and $700\,$MeV (blue curve) in symmetric nuclear matter. }
\label{fig:Nmass}
\end{figure}

As in  Ref.~\cite{Harada:2016uca},
we define the effective nucleon masses as the energies at ${\bf k}=0$:
\begin{align}
m_n^{\rm(eff)} &= m_N + g_{\omega NN} \bar\omega -g_{\rho NN}\bar\rho  \notag\\
m_p^{\rm(eff)} &= m_N + g_{\omega NN}\bar\omega + g_{\rho NN}\bar\rho \ .
\label{eff mass N}
\end{align}
In Fig.~\ref{fig:Nmass}, we plot the density dependence of these effective masses
in symmetric matter, which are given by taking $\bar\rho=0$ in Eq.~(\ref{eff mass N}),
for $m_{N0}=500$ and $700\,$MeV.
 This shows that the effective mass $m_{N}^{\rm(eff)}$ for $m_{N0}=500$\,MeV is quite similar to the one for $m_{N0}=700$\,MeV for $\rho_B \lesssim \rho_0$.  A reason is as follows:
As in Ref.~\cite{Motohiro:2015}, we use the saturation density $\rho_0 = 0.16\,\mbox{fm}^{-3}$ and $\mu_B = 923\,\mbox{MeV}$ at saturation density as inputs.
The former leads to the Fermi momentum of nucleons as $k_F \simeq 270\,\mbox{MeV}$.
Then, from Eqs.~(\ref{N energy}) and (\ref{eff mass N}), we obtain
\begin{align}
\mu_B - m^{(eff)}_N = & \sqrt{ k_F^2 + m_N^2 } - m_N^2 \simeq \frac{ k_F^2}{2 m_N}
\notag\\
\simeq & \frac{ k_F^2}{ 2 \mu_B} \simeq 40 \,\mbox{MeV} \ ,
\end{align}
independently of the choice of the chiral invariant mass $m_{N0}$.

In the density region higher than $\rho_0$, Fig.~\ref{fig:Nmass} shows that $m_N^{\rm (eff)}$ decreases against increasing density for $m_{N0}=500$\,MeV
since $m_N$ in Eq.~(\ref{N mass}) decreases rapidly.
On the other hand, $m_N^{\rm (eff)}$ for $m_{N0}=700$\,MeV increases
since $m_N$ in Eq.~(\ref{N mass}) decreases very slowly and the large $\omega$ contribution pushes up the effective mass.

The mass of the $\Delta$ baryon with positive parity is obtained as
\begin{align}
m_\Delta&=\frac{1}{2}\left(\sqrt{(a+b)^2\bar\sigma^2+4m^2_{\Delta0}} - (a-b)\bar\sigma\right) \ ,
\end{align}
where $m_{\Delta0}$ is the chiral invariant mass of $\Delta$.
The mass of the $\Delta$ baryon with  negative parity is given by
\begin{align}
m_{\Delta^\ast}&=\frac{1}{2}\left(\sqrt{(a+b)^2\bar\sigma^2+4m^2_{\Delta0}} + (a-b)\bar\sigma\right) \ .
\end{align}
Using the masses of $\Delta(1232)$ and $\Delta(1700)$ in free space as inputs, we determine the values of the couplings $a$ and $b$ for fixed $m_{\Delta0}$,
which is summarized in Table~\ref{table:ab} for typical values of $m_{\Delta 0}$.
		\begin{table}[htbp]
			\caption{Values of $\sigma\Delta\Delta$ coupling constants for several choices of the chiral invariant mass $m_{\Delta0}$. }
				\begin{tabular}{cccccc}
					\hline \hline
						$m_{^\Delta 0}$&$500$ & $700$&$1000$&1300&1400\\
					\hline
						$a$ &17.5&16.5&14.1&9.87&7.25\\
						$b$ &12.2&11.4&9.08&4.87&2.18\\
 					\hline \hline
				\label{table:ab}
 			\end{tabular}
		\end{table}

The dispersion relations for the $\Delta$ baryons with positive parity in nuclear matter are given by
\begin{align}
E_{\Delta^{++}} = & \sqrt{m_\Delta^2+{\bf k}^2}+g_{\omega \Delta\Delta}\bar\omega + 3 g_{\rho \Delta\Delta}\bar\rho \ ,
\notag\\
E_{\Delta^{+}} = & \sqrt{m_\Delta^2+{\bf k}^2}+g_{\omega \Delta\Delta}\bar\omega +  g_{\rho \Delta\Delta}\bar\rho \ ,
\notag\\
E_{\Delta^{0}} = & \sqrt{m_\Delta^2+{\bf k}^2}+g_{\omega \Delta\Delta}\bar\omega  - g_{\rho \Delta\Delta}\bar\rho \ ,
\notag\\
E_{\Delta^{-}} = & \sqrt{m_\Delta^2+{\bf k}^2}+g_{\omega \Delta\Delta}\bar\omega - 3 g_{\rho \Delta\Delta}\bar\rho \ ,
\label{Delta energy}
\end{align}
where
$g_{\omega\Delta\Delta}$ and $g_{\rho\Delta\Delta}$ are the couplings of the $\omega$ and $\rho$ mesons to the $\Delta$ baryon,
and the
energies are measured from their respective chemical potentials given by
\begin{align}
\mu_{\Delta^{++} } & = \mu_B + \frac{3}{2} \mu_I \ , \notag\\
\mu_{\Delta^+} & = \mu_B + \frac{1}{2} \mu_I \ , \notag\\
\mu_{\Delta^0 } & = \mu_B - \frac{1}{2} \mu_I \ , \notag\\
\mu_{\Delta^- } & = \mu_B - \frac{3}{2} \mu_I \, .
\label{Delta chemical}
\end{align}

\begin{figure}[htbp]
\vspace{-85pt}
\includegraphics[bb=0 0 480 360,width=10cm]{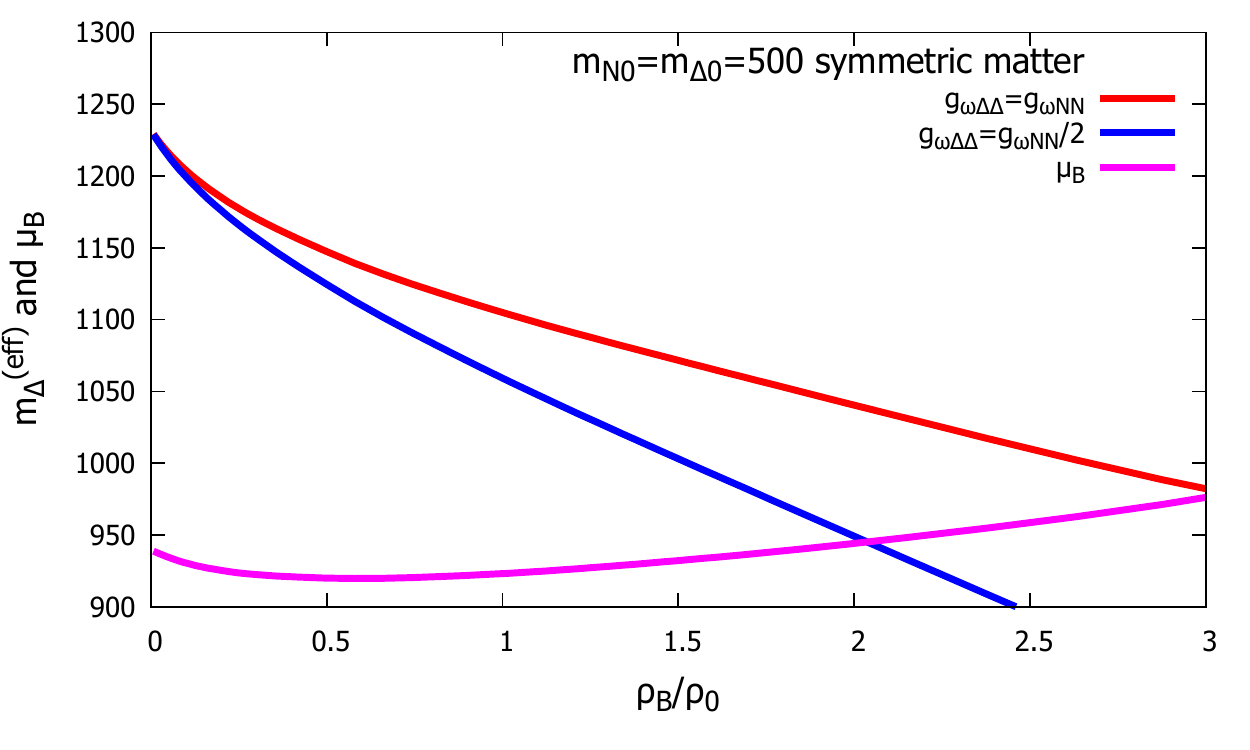}
\caption{Density dependence of the effective masses of $\Delta$ for $g_{\omega \Delta\Delta} = g_{\omega NN}$ (red curve) and $g_{\omega \Delta\Delta} = g_{\omega NN}/2$ (blue curve) with fixed values of $m_{N0}=m_{\Delta 0}=500\,$MeV in symmetric nuclear matter.
The pink curve shows the baryon chemical potential $\mu_B$.
}
\label{fig:delmass500}
\end{figure}

\begin{figure}[htbp]
\vspace{-85pt}
\includegraphics[bb=0 0 480 360,width=10cm]{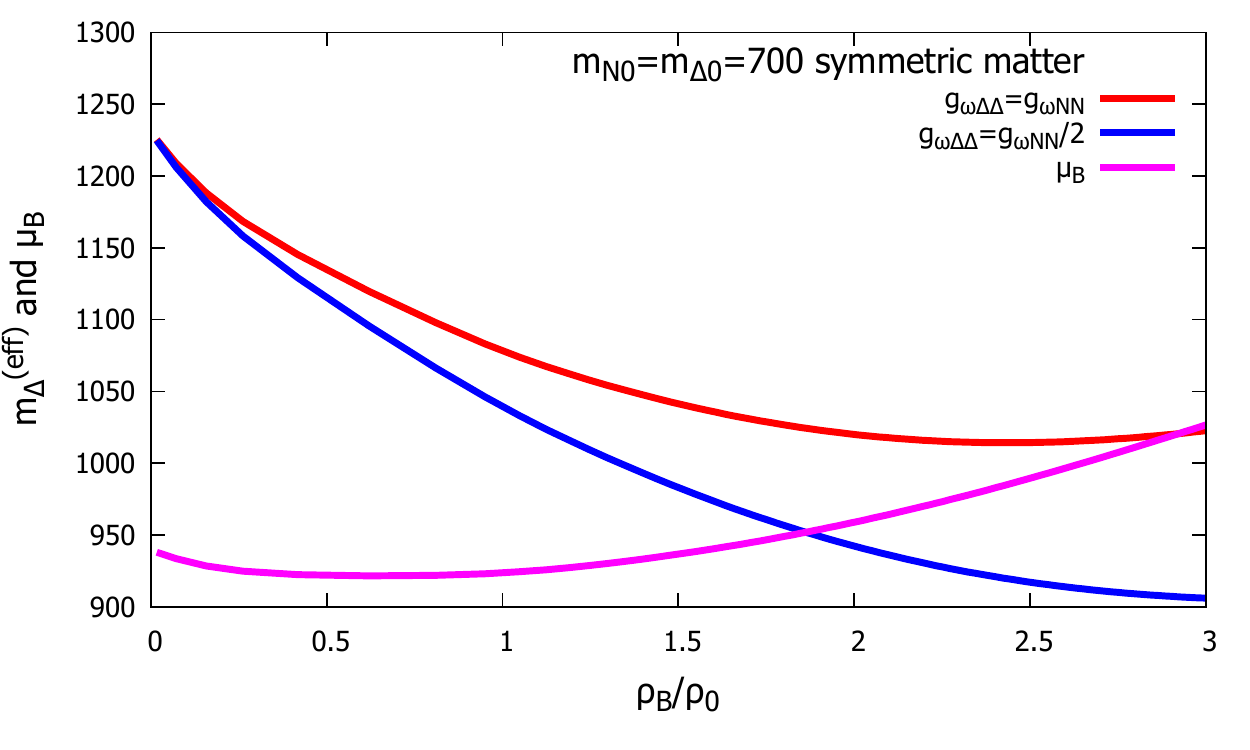}
\caption{Density dependence of the effective masses of $\Delta$ for $g_{\omega \Delta\Delta} = g_{\omega NN}$ (red curve) and $g_{\omega \Delta\Delta} = g_{\omega NN}/2$ (blue curve) with fixed values of $m_{N0}=m_{\Delta 0}=700\,$MeV in symmetric nuclear matter.
The pink curve shows the baryon chemical potential $\mu_B$.
}\label{fig:delmass700}
\end{figure}

We define the effective masses as
\begin{align}
m_{\Delta^{++}}^{\rm(eff)} = & m_\Delta+g_{\omega \Delta\Delta}\bar\omega + 3 g_{\rho \Delta\Delta}\bar\rho \ ,
\notag\\
m_{\Delta^{+}}^{\rm(eff)} = & m_\Delta+g_{\omega \Delta\Delta}\bar\omega +  g_{\rho \Delta\Delta}\bar\rho \ ,
\notag\\
m_{\Delta^{0}}^{\rm(eff)} = & m_\Delta+g_{\omega \Delta\Delta}\bar\omega  - g_{\rho \Delta\Delta}\bar\rho \ ,
\notag\\
m_{\Delta^{-}}^{\rm(eff)} = & m_\Delta+g_{\omega \Delta\Delta}\bar\omega - 3 g_{\rho \Delta\Delta}\bar\rho \, .
\label{Delta eff mass}
\end{align}
We plot the density dependence of the effective mass in symmetric matter ($\bar\rho=0$) for
$m_{\Delta 0} = 500$ and $700$\,MeV
together with the baryon chemical potential $\mu_B$ in Figs.~\ref{fig:delmass500} and \ref{fig:delmass700}.
These figures show that the effective mass of $\Delta$ becomes smaller than the baryon chemical potential around
$\rho_B / \rho_0 \sim 2 $-$3$, which indicates that
the $\Delta$ baryon is populated in the ground state, forming $\Delta$ fermi sea.

\section{Delta matter}\label{deltaM}

We now construct a thermodynamical potential with the $\Delta$ baryons following Ref.~\cite{Motohiro:2015}:
\begin{align}
\Omega = &-\frac{1}{2}m_\rho^2\bar{\rho}^2-\frac{1}{2}m_\omega^2\bar{\omega}^2 + V_\sigma \notag\\
& - \sum_{\alpha = p,n} \int \frac{dk}{\pi^2} k^2 ( \mu_\alpha - E_\alpha ) \, \theta( \mu_\alpha - E_\alpha ) \notag\\
& - 2 \sum_{a = {++},+,0,-} \int \frac{dk}{\pi^2} k^2 ( \mu_{\Delta^a} - E_{\Delta^a} ) \, \theta( \mu_{\Delta^a} - E_{\Delta^a} ) \notag \\
&+({\rm negative \ parity }) ,  \label{thermodynamic potential} \\
& V_\sigma = \frac{1}{2}\bar\mu^2\bar{\sigma}^2-\frac{1}{4}\lambda_4\bar{\sigma}^4+\frac{1} {6}\lambda_6\bar{\sigma}^6-\bar m\epsilon\bar{\sigma} \ ,
\label{V sigma}
\end{align}
where $E_{p,n}$ and $E_{\Delta^{++},\Delta^+,\Delta^0,\Delta^-}$ are the effective energies given in
Eqs.~(\ref{N energy}) and (\ref{Delta energy}), $\mu_{p,n}$ and $\mu_{\Delta^{++},\Delta^+,\Delta^0,\Delta^-}$ are the chemical potentials given in Eqs.~(\ref{N chemical}) and (\ref{Delta chemical}).
In this expression, we drop the antiparticles because they do not contribute at zero temperature.

The values of the mean fields $\bar{\sigma}$, $\bar{\omega}$ and $\bar{\rho}$ are determined by the stationary conditions for the above thermodynamical potential.
The stationary condition for $\bar{\omega}$ is written as
\begin{align}
m_\omega^2 \bar{\omega} = & g_{\omega NN} \sum_{\alpha = p,n} \rho_\alpha + g_{\omega \Delta\Delta} \sum_{a = {++},+,0,-} {\rho}_{_{\Delta^a}} \notag\\
& +({\rm negative \ parity }) \ ,
\label{min omega}
\end{align}
where
$\rho_\alpha$ and $\rho_{_{\Delta^a}}$ are the number densities of the corresponding baryons
\begin{align}
\rho_\alpha = & \int \frac{dk}{\pi^2} k^2 \, \theta( \mu_\alpha - E_\alpha ) \ , \\
\rho_{_{\Delta^a}} = & 2 \int \frac{dk}{\pi^2} k^2 \, \theta( \mu_{\Delta^a} - E_{\Delta^a} )  \ .
\end{align}
The stationary condition for $\bar{\rho}$ is expressed as
\begin{align}
m_\rho^2 \bar{\rho} = & g_{\rho NN} \left( \rho_p - \rho_n \right) \notag\\
& + g_{\rho \Delta\Delta} \left( 3 \rho_{_{\Delta^{++}}} + \rho_{_{\Delta^+}} - \rho_{_{\Delta^0}} - 3 \rho_{_{\Delta^-}} \right) \notag\\
& +({\rm negative \ parity }) \ .
\label{min rho}
\end{align}
The stationary condition for $\bar{\sigma}$ is expressed as
\begin{align}
 &
\sum_{\alpha = p , n} \, \int \frac{dk}{\pi^2} \, \frac{ k^2 m_N }{ \sqrt{ m_N^2 + k^2 } } \frac{ \partial m_N }{ \partial \bar{\sigma} } \, \theta( \mu_\alpha - E_\alpha )  \notag\\
& +
2 \sum_{a = {++},+,0,-} \int \frac{dk}{\pi^2} \, \frac{ k^2 m_\Delta }{ \sqrt{ m_\Delta^2 + k^2 } } \frac{ \partial m_\Delta }{ \partial \bar{\sigma} } \, \theta( \mu_{\Delta^a} - E_{\Delta^a} ) \notag\\
& +({\rm negative \ parity }) \notag\\
& + \frac{\partial V_\sigma}{\partial \bar{\sigma} } = 0 \ .
\label{min sigma}
\end{align}

There are sixteen parameters in the present model:
\begin{align}
& m_\rho \ , \quad m_\omega \ , \quad \bar{\mu} \ ,\quad \lambda_4 \ , \quad \lambda_6 \ , \quad \bar{m} \epsilon \ , \notag\\
& m_{N0} \ , \quad g_{\sigma N1} \ , \quad g_{\sigma N2} \ , \quad g_{\omega NN} \ , \quad g_{\rho NN} \ , \notag\\
& m_{\Delta 0} \ , \quad g_{\sigma\Delta 1} \ , \quad g_{\sigma\Delta 2} \ , \quad g_{\omega \Delta\Delta} \ , \quad g_{\rho \Delta\Delta} \ .
\end{align}
Following Ref.~\cite{Motohiro:2015}, we use the masses of $\omega$ and $\rho$ mesons, pions, $N(939)$ and $N(1535)$ as well as the pion decay constant in vacuum as inputs.
We also use the masses of $\Delta(1232)$ and $\Delta(1700)$ as inputs.
We list the values of the physical inputs in vacuum in Table~\ref{vacinput}.
\begin{table}[htbp]
\begin{center}
\caption{Physical inputs in vacuum (MeV).}\label{vacinput}
\begin{tabular}{c|c|c|c|c|c|c|c}
\hline
$m_{N+}$ & $m_{N-}$ & $m_{\Delta +}$ & $m_{\Delta-}$ &$m_\omega$ & $m_\rho$ & $f_\pi$ &$m_\pi$ \\
\hline\hline
 939 & 1535 & 1232 & 1700 & 783 & 776 & $92.3$ & $140$ \\
\hline\hline
\end{tabular}
\end{center}
\end{table}
In addition, we use the nuclear matter saturation density, the binding energy, the incompressibility and the symmetry energy at the normal nuclear matter density as inputs.
We list the empirical values of them, which we use as inputs, in Table~\ref{mat-input}.
\begin{table}[htbp]
\begin{center}
\caption{Physical inputs at the normal nuclear matter density.
$\rho_0$: saturation density,
$E_{\rm bind}$: binding energy, $K$: incompressibility, $E_{\rm sym}$: symmetry energy.
}
\label{mat-input}
\begin{tabular}{c|c|c|c}
\hline
$\rho_0$ & $E_{\rm bind}$ & $K$ & $E_{\rm sym}$ \\
\hline\hline
$0.16$\,fm$^{-3}$ & $-16$\,MeV & $240$\,MeV & $31$\,MeV \\
\hline\hline
\end{tabular}
\end{center}
\end{table}

Altogether we have only twelve inputs, so we regard $m_{N0}$, $m_{\Delta0}$, $g_{\omega \Delta\Delta}$ and $g_{\rho \Delta\Delta}$ as free parameters, and show our results for given values of these four parameters.
Below we shall briefly sketch how to fix the other  twelve parameters.

The values of $m_\omega$ and $m_\rho$ are trivially determined by Table~\ref{vacinput}.
The value of $\bar{m}\epsilon$ is fixed as
\begin{equation}
\bar{m}\epsilon = m_\pi^2 f_\pi \ ,
\end{equation}
from the values in Table~\ref{vacinput}.
The values of the four couplings of $\sigma$ to $N$ and $\Delta$ are fixed as in Tables~\ref{table:sigmaNN} and \ref{table:ab}.
The value of $\bar{\mu}$ is determined once the values of $\lambda_4$ and $\lambda_6$ are determined via the stationary condition of $V_\sigma$ in Eq.~(\ref{V sigma}) in vacuum:
\begin{equation}
\bar{\mu}^2 = \lambda_4 f_\pi^3 - \lambda_6 f_\pi^5 - \bar{m} \epsilon \ .
\label{det mu}
\end{equation}
Imposing that the energy density is minimized at $\rho_0$, we obtain  the pressure  vanishing  at $\rho_0$:
\begin{equation}
P \big \vert _{\mu_B = \mu_{B0}, \mu_I = 0}  = - \Omega \big \vert _{\mu_B = \mu_{B0}, \mu_I = 0}   = 0 \ .
\label{mat inputs form 0}
\end{equation}
Then,
using $E_{\rm bind} = - 16$\,MeV at $\rho_B = \rho_0$ 
and $m_{N+} = 939$\,MeV at vacuum,
 we get $\mu_B = \mu_{B0} = 923$\,MeV:
\begin{equation}
\rho_0 =  \rho_B \big \vert _{\mu_B = \mu_{B0}, \mu_I = 0} = \left(\frac{\partial\Omega}{\partial\mu_B}\right) _{\mu_B = \mu_{B0}, \mu_I = 0}
\ .
\label{mat inputs form 1}
\end{equation}
In addition to these, the incompressibility $K$ and the symmetry energy $E_{\rm sym}$ at the normal nuclear matter density are determined by
\begin{align}
K =& 9\rho_0\frac{\partial\mu_B}{\partial\rho_B} \bigg \vert _{\mu_B = \mu_{B0}, \mu_I = 0}\ , \notag\\
E_{\rm sym} =& 4\rho_0\frac{\partial\mu_I}{\partial\rho_I}  \bigg \vert _{\mu_B = \mu_{B0}, \mu_I = 0} \, .
\label{mat inputs form 2}
\end{align}
First,
for given values of $m_{N0}$, $m_{\Delta0}$ and the ratio $g_{\omega \Delta\Delta}/g_{\omega NN}$,  we determine the values of $\bar{\omega}$ and $g_{\omega NN}$ using $\rho_0$ in Eq.~(\ref{mat inputs form 1})  and the stationary condition for $\bar{\omega}$ in Eq.~(\ref{min omega}).
Second,   we fix the values of $\lambda_4$, $\lambda_6$ and $\bar\sigma$ using the pressure $P$ in  Eq.~(\ref{mat inputs form 0}),
$K$ in Eq.~(\ref{mat inputs form 2}) and the stationary condition for $\bar{\sigma}$ in Eq.~(\ref{min sigma}).
Third, for a given value of $g_{\rho\Delta\Delta}/g_{\rho NN}$ we determine the values of $\bar{\rho}$ and $g_{\rho NN}$ from $E_{\rm sym}$ in Eq.~(\ref{mat inputs form 2}) and the stationary condition for $\bar{\rho}$ in Eq.~(\ref{min rho}).
Once we fix the parameters as above, we can calculate several physical quantities with given chemical potentials $\mu_B$ and $\mu_I$.

\begin{figure}[htbp]
\begin{center}
\vspace{-85pt}
\includegraphics[bb=0 0 480 360,width=10cm]{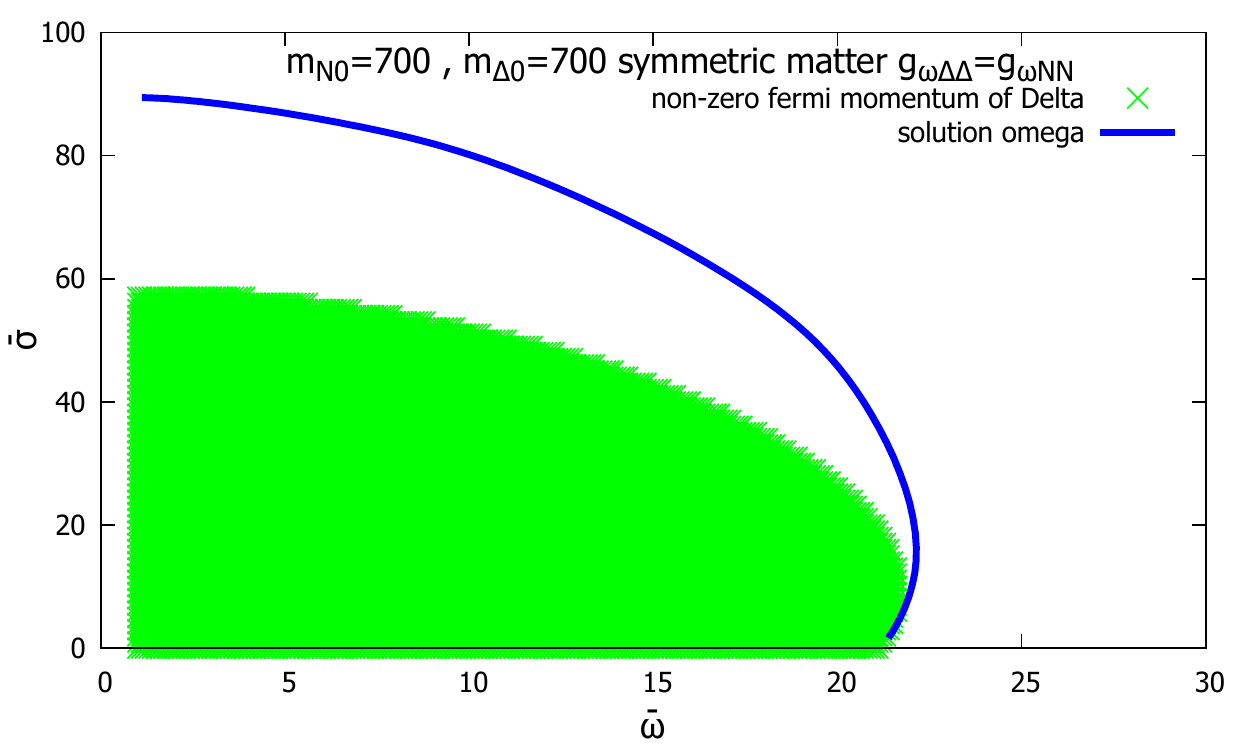}
\end{center}
\caption{
Relations between $\bar{\omega}$ and $\bar{\sigma}$ for $m_{N0}= m_{\Delta 0} = 700$\,MeV and $g_{\omega \Delta\Delta} = g_{\omega NN}$.
The blue-colored area is determined from the relation in Eq.~(\ref{omega rho})
  and the stationary condition for $\omega$ in Eq.~(\ref{min omega}), while the green-colored area is settled by requiring $\mu_{\Delta} - E_{\Delta} > 0$.
}
\label{sigmaomega}
\end{figure}

Let us first show that the $\Delta$ baryons are unlikely to exist in  nuclear matter with $\rho_B \sim \rho_0$
for $m_{N0}= m_{\Delta 0} = 700$\,MeV and $g_{\omega \Delta\Delta} = g_{\omega NN}$.
From the stationary condition for $\bar{\omega}$ at $\rho_B = \rho_0$ with $g_{\omega \Delta\Delta} = g_{\omega NN}$, we obtain
\begin{equation}
m_\omega^2 \bar{\omega} = g_{\omega NN} \rho_0 \ ,
\label{omega rho}
\end{equation}
which fixes the value of $g_{\omega NN}$ for given value of $\bar{\omega}$.
Then, we determine the the value
of $\bar{\sigma}$ using Eq.~(\ref{min omega}), which is shown by the blue curve
in Fig.~\ref{sigmaomega}.
On the other hand, for given values of $\bar{\omega}$ and $\bar{\sigma}$ we check if $\mu_{\Delta} - E_{\Delta} > 0$ ($\mu_I = \bar{\rho}=0$) is satisfied.
The green-colored area in Fig.~\ref{sigmaomega} shows the region where $\mu_{\Delta} - E_{\Delta} > 0$ is satisfied.
Since the value of $\bar{\sigma}$ cannot be so small, from this figure, we conclude that the $\Delta$ baryon cannot enter the matter at the normal nuclear matter density for this parameter choice.
In the following, we will show our results with several parameter choices.
For all cases we have checked that the $\Delta$ baryon does not pile up in the ground state, i.e., no transition to $\Delta$ matter, at or near the normal nuclear matter density.

\begin{figure}[htbp]
\begin{center}
\vspace{-85pt}
\includegraphics[bb=0 0 480 360,width=10cm]{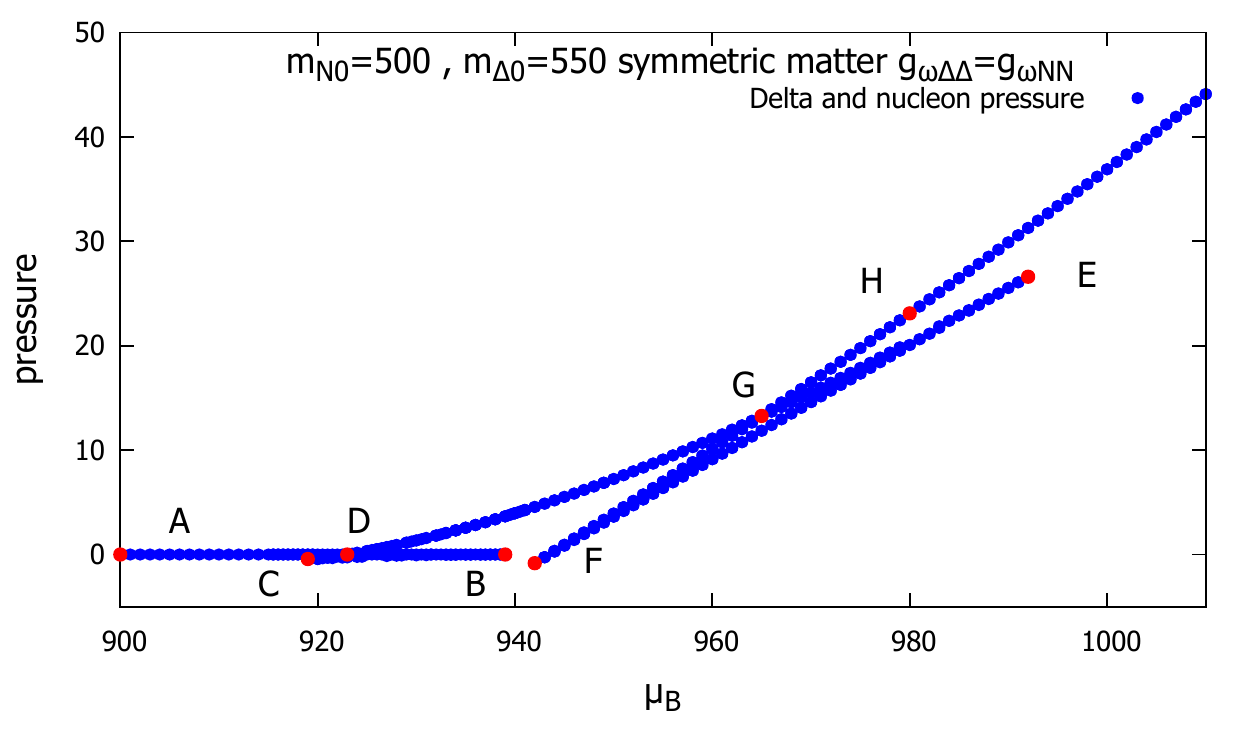}
\end{center}
\caption{Relation between the chemical potential  (horizontal axis) and the pressure (vertical axis)
for a parameter choice of
$m_{N0}=500$\,MeV, $m_{\Delta0}=550$\,MeV
and $g_{\omega \Delta\Delta} = g_{\omega NN}$.
Unit of $\mu_B$ is MeV and the one of pressure is MeV$\cdot$fm$^{-3}$. Explanation for the points A-H is given in the text. }
\label{P muB 500 550}
\end{figure}

\begin{figure}[htbp]
\begin{center}
\vspace{-85pt}
\includegraphics[bb=0 0 480 360,width=10cm]{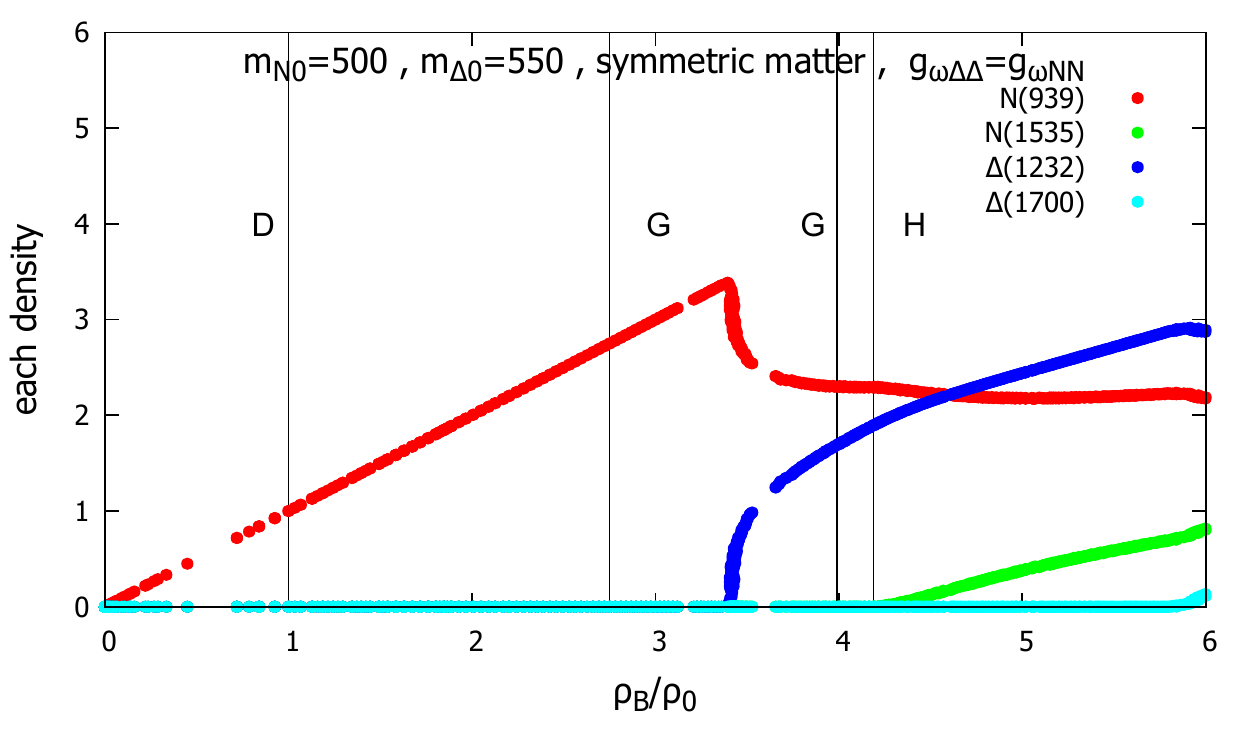}
\end{center}
\caption{
Densities of $N(939)$, $\Delta(1232)$ and $N^\ast(1535)$.
  Horizontal axis shows the baryon number density scaled by the normal nuclear matter density $\rho_0$.  Vertical axis shows densities of
  $N(939)$ (red curve),
  $\Delta(1232)$ (blue curve)
and $N^\ast(1535)$ (green curve)
scaled by $\rho_0$.
}
\label{density 500 550}
\end{figure}

\begin{figure}[htbp]
\begin{center}
\vspace{-85pt}
\includegraphics[bb=0 0 480 360,width=10cm]{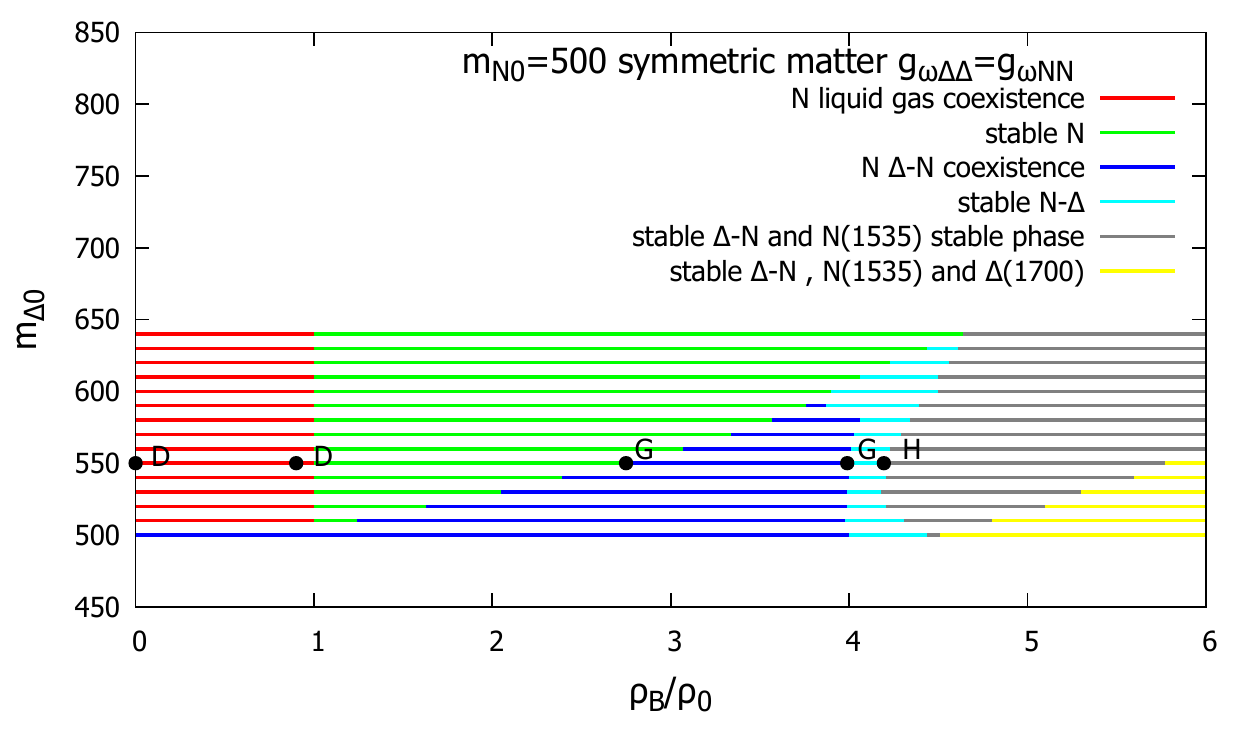}
\end{center}
\caption{
Summary of matter constituents for $m_{N0} = 500$\,MeV
and $g_{\omega \Delta\Delta} = g_{\omega NN}$ in symmetric nuclear matter.  The horizontal axis shows the baryon number density scaled  by the normal nuclear matter density, and the vertical axis shows the value of  $m_{\Delta 0}$ in unit of MeV.
The red area indicates the ``$N$ liquid-gas coexistence'',
the green  the  ``stable $N$ matter'',
the blue  the  ``coexistence of $\Delta$-$N$ matter and $N$ matter'',
the cyan  the ``stable $N$-$\Delta$ matter''.
In the black area, $N^\ast(1535)$ enters into matter, while in the yellow area, both $N^\ast(1535)$ and $\Delta(1700)$ enter.
}
\label{phase 500 1}
\end{figure}

We now show our results for symmetric matter ($\mu_I = 0$).
In Fig.~\ref{P muB 500 550}, for a parameter choice of
$m_{N0}=500$\,MeV, $m_{\Delta0}=550$\,MeV
and $g_{\omega \Delta\Delta} = g_{\omega NN}$, we show the resultant relation between the chemical potential $\mu_B$ and the pressure.
Since the stationary condition for $\bar{\sigma}$ is non-linear, we often have a few solutions for a given $\mu_B$ with fixed parameters.  Then, there exist a few values of pressure for a given $\mu_B$.
On the strait line AB except the point D, we have a solution corresponding to vacuum, $\bar{\sigma}=f_\pi$ and $\bar{\omega}=0$, so that the pressure and the density are zero, $P=0$ and $\rho_B = 0$.
At the point D, we have two solutions: one corresponds to the vacuum and another to normal nuclear matter.
Along the curve from the point B to D through C, there exists a solution corresponding to the negative pressure with the density smaller than the normal nuclear matter density.
In Fig.~\ref{density 500 550}, we show the densities of
$N(939)$, $\Delta(1232)$ and $N(1535)$
against the baryon number density $\rho_B$ given as sum of these
densities.
This figure shows that only $N(939)$ exists for $\rho_B < \rho_0$.
Since the pressure is negative, we call this region the  ``$N$ liquid-gas coexistence'',
which is indicated by the red area in Fig.~\ref{phase 500 1}.

When the pressure increases along the curve from the point D to G in Fig.~\ref{P muB 500 550},  the nucleon density increases from $\rho_B/\rho_0 = 1 $ to
 about $2.7$
as shown in Fig.~\ref{density 500 550}.  Here the ordinary nuclear matter exists and we call this region the ``stable $N$ matter''
indicated by the green area in Fig.~\ref{phase 500 1}.

Along the curve from G to E, there is a solution in which the nucleon density keeps increasing,  which corresponds to
$\rho_B/\rho_0 \simeq 2.7$ to $3.4$
in Fig.~\ref{density 500 550}.
In this region we have another solution having larger pressure for the same value of the chemical potential along the curve GH.  Then, the region between G and E is not energetically favored.
Along the curve from E to  G through F,  the $\Delta$ baryon appears in matter as in Fig.~\ref{density 500 550}.  This region is also unfavored
 since we have another solution having the larger pressure for the same value of the chemical potential along DG.  We call this region
  $\rho_B/\rho_0 \simeq 2.7$ - $4$
 the ``coexistence of $\Delta$-$N$ matter and nuclear matter''
 indicated by the blue area in Fig.~\ref{phase 500 1}.

At the point G in Fig.~\ref{P muB 500 550}, there is a solution corresponding to
 $\rho_B/\rho_0 \simeq 4$,
at which  both the nucleon and $\Delta$ exist in matter as shown in Fig.~\ref{density 500 550}.
At $\rho_B/\rho_0 \simeq 4.2$ which corresponds to the point H in Fig.~\ref{P muB 500 550},
$N^\ast(1535)$ also enters into the matter as shown in Fig.~\ref{density 500 550}.
Between the points G and H,
there is a stable matter including the $\Delta$ baryon and the nucleon, which we call the
 ``stable $N$-$\Delta$ natter''
indicated by cyan region in Fig.~\ref{phase 500 1}.~\footnote{One may wonder why we have more $\Delta$ baryons than nucleons in stable $N$-$\Delta$  matter in Fig.~\ref{density 500 550}.
In this $N$-$\Delta$  matter,
the effective masses of $\Delta$ and $N$ are close to each other as in
Eq.~(\ref{ef masses}).  Since $\Delta$ has sixteen degrees of freedom (spin $3/2$ and isospin $3/2$) compared with four (spin $1/2$ and isospin $1/2$) for nucleon, the $\Delta$ density could be larger than $N$ density.}

In Fig.~\ref{phase 500 1} we provide  a summary of
 matter constituents
on $\rho_B/\rho_0$-$m_{\Delta 0}$ plane with
 $m_{N0} = 500$\,MeV
and $g_{\omega \Delta\Delta} = g_{\omega NN}$.
This shows that, for  $m_{\Delta 0} = 500$\,MeV,
the
``coexistence of $\Delta$-$N$ matter and nuclear matter''
is realized in  the wide density region for
 $\rho_B/\rho_0 <4$.
This implies the non-existence of  normal nuclear matter, and so this parameter choice is excluded.
For $m_{\Delta 0}> 500$\,MeV,
on the other hand, there exists the liquid-gas phase transition of ordinary nuclear matter at $\rho_B/\rho_0 = 1$ from the  ``$N$ liquid-gas coexistence''
indicated by red area to the  ``stable $N$ matter''
by the green area.
For $m_{\Delta 0} = 510$\,MeV, the  ``coexistence of $\Delta$-$N$ matter and nuclear matter'' starts at $\rho_B/\rho_0 \simeq 1.2$. The onset density of the appearance of $\Delta$ increases with increasing $m_{\Delta 0}$ for $m_{\Delta 0} \le 590$\,MeV.
Since the larger $m_{\Delta 0}$ leads to smaller $\sigma \Delta\Delta$ couplings as shown in
Table~\ref{table:ab}, this tendency is consistent with the ones obtained in
Refs.~\cite{Li:1997yh,Lavagno:2012bn}.

For $m_{\Delta 0} \le 590$\,MeV, the ``stable $N$-$\Delta$ matter'' indicated by cyan area in Fig.~\ref{phase 500 1} appears at $\rho_B/\rho_0 \sim 4$, independently of the value of $m_{\Delta0}$.

This implies that the first order phase transition occurs for  $510 \le m_{\Delta 0} \le 590$,
 and that, for example, the density jumps from $\rho_B/\rho_0 \simeq 1.2$ to  $4$ for $m_{\Delta 0} = 510$\,MeV.

For $m_{\Delta 0} \ge 600$\,MeV, the  ``stable $N$ matter'' by green area is smoothly connected to the ``stable $N$-$\Delta$ matter'' indicated by cyan area.
Furthermore, for $\rho_B/\rho_0 \sim 4.2$ - $4.5$, $N^\ast(1535)$ enters into matter, which is a reflection of the partial restoration of chiral symmetry accelerated by $\Delta$ matter.  We will investigate this point in detail in the next section.

\begin{figure}[htbp]
\begin{center}
 (a) \ \ \\
\vspace{-85pt}
\includegraphics[bb=0 0 480 360,width=10cm]{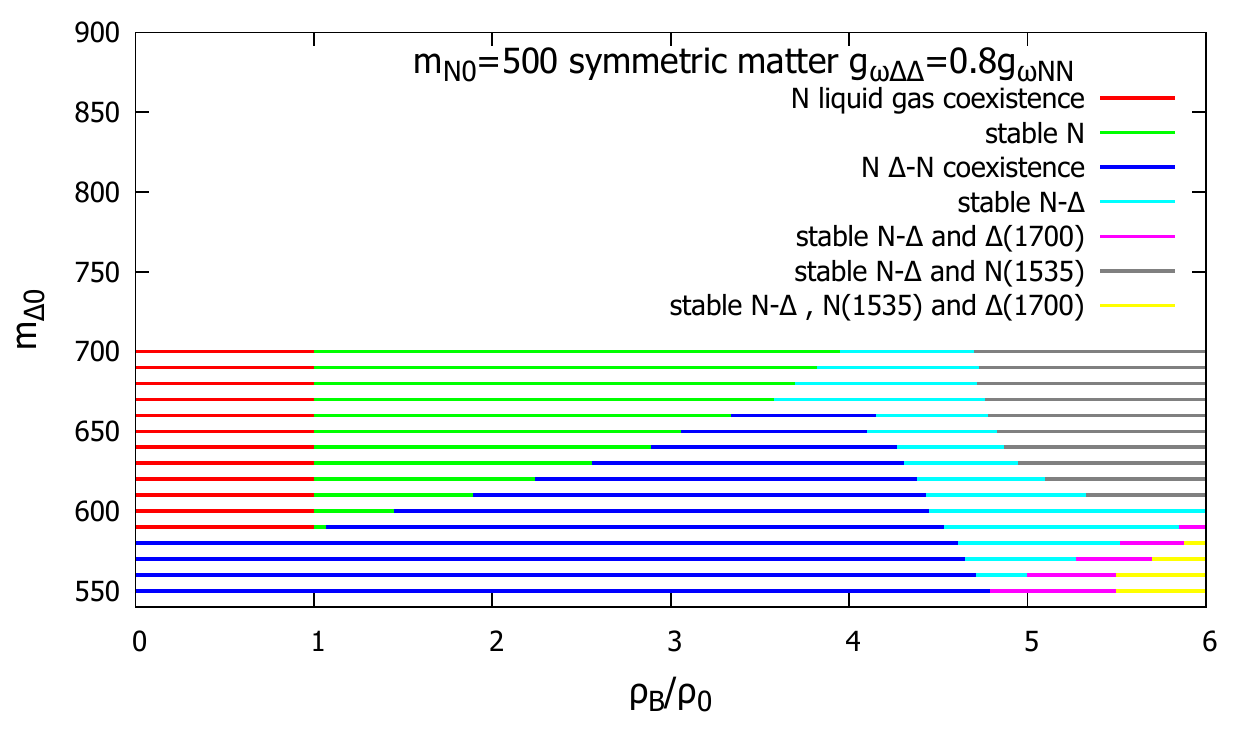}\\
 (b) \ \   \\
\vspace{-85pt}
\includegraphics[bb=0 0 480 360,width=10cm]{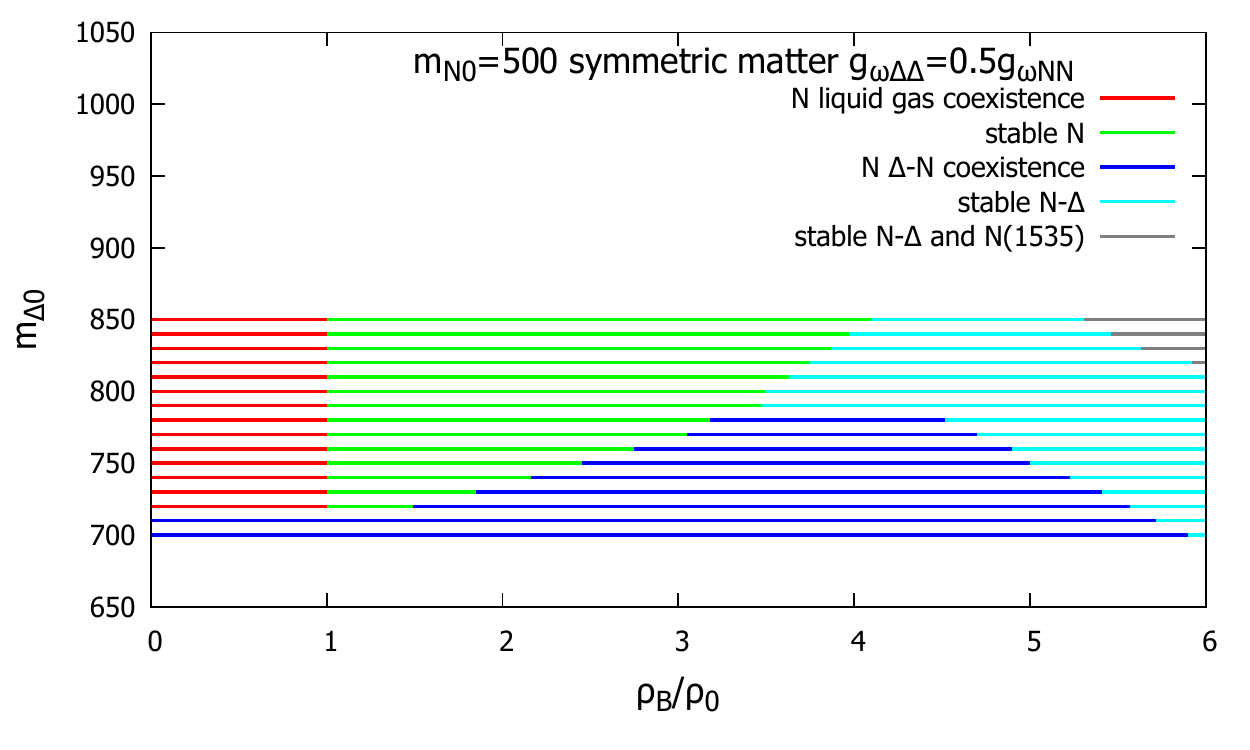}\\
\end{center}
\caption{
Matter constituents
for (a)~$g_{\omega \Delta\Delta} = 0.8 g_{\omega NN}$  and (b)~$g_{\omega \Delta\Delta} = 0.5 g_{\omega NN}$
 with fixed value of   $m_{N 0} = 500$\,MeV
 in symmetric matter.
The horizontal axis shows the baryon number density scaled  by the normal nuclear matter density, and the vertical axis shows the value of  $m_{\Delta 0}$.
The red area indicates the  ``$N$ liquid-gas coexistence'',
the green area the  ``stable $N$ matter'',
the blue area the  ``coexistence of $\Delta$-$N$ matter and nuclear matter'',
the cyan area the  ``stable $N$-$\Delta$ matter''.
In the black, pink and yellow areas, $N^\ast(1535)$, $\Delta(1700)$, and both of them enter into matter, respectively.
}
\label{phase 500 2}
\end{figure}

In Fig.~\ref{phase 500 2}, we show
matter constituents
for different choices of $g_{\omega \Delta\Delta}$ with fixed $m_{N 0} = 500$\,MeV.

This shows that the following qualitative structures are similar:
(1) The onset density of the ``coexistence of $\Delta$-$N$ matter and nuclear matter'' is larger for larger value of $m_{\Delta0}$.  (2) There exists the ``stable $N$-$\Delta$ matter''.

There are several points which depend on the value of $g_{\omega \Delta\Delta}$:
(1) The minimum value of $m_{\Delta 0}$, for which
the onset density of $\Delta$ matter
is larger than the normal nuclear matter density,  is larger
for smaller $g_{\omega \Delta\Delta}$; $m_{\Delta 0}^{\rm(min)} = 510$, $590$ and $720$\,MeV for $g_{\omega\Delta\Delta}/g_{\omega NN} = 1$, $0.8$, $0.5$, respectively.
This feature is consistent with the one obtained in Ref.~\cite{Li:1997yh}.
(2) For some values of $m_{\Delta0}$, in high density region, there appear $N^\ast(1535)$ and/or $\Delta(1700)$, which are chiral partners to $N(929)$ and $\Delta(1232)$, reflecting the partial chiral symmetry restoration.  The onset density of the appearance is larger for smaller $g_{\omega \Delta\Delta}$.
We will investigate this point further in the next section.

\begin{figure}[htbp]
\begin{center}
 (a) \ \  \\
\vspace{-85pt}
\includegraphics[bb=0 0 480 360,width=10cm]{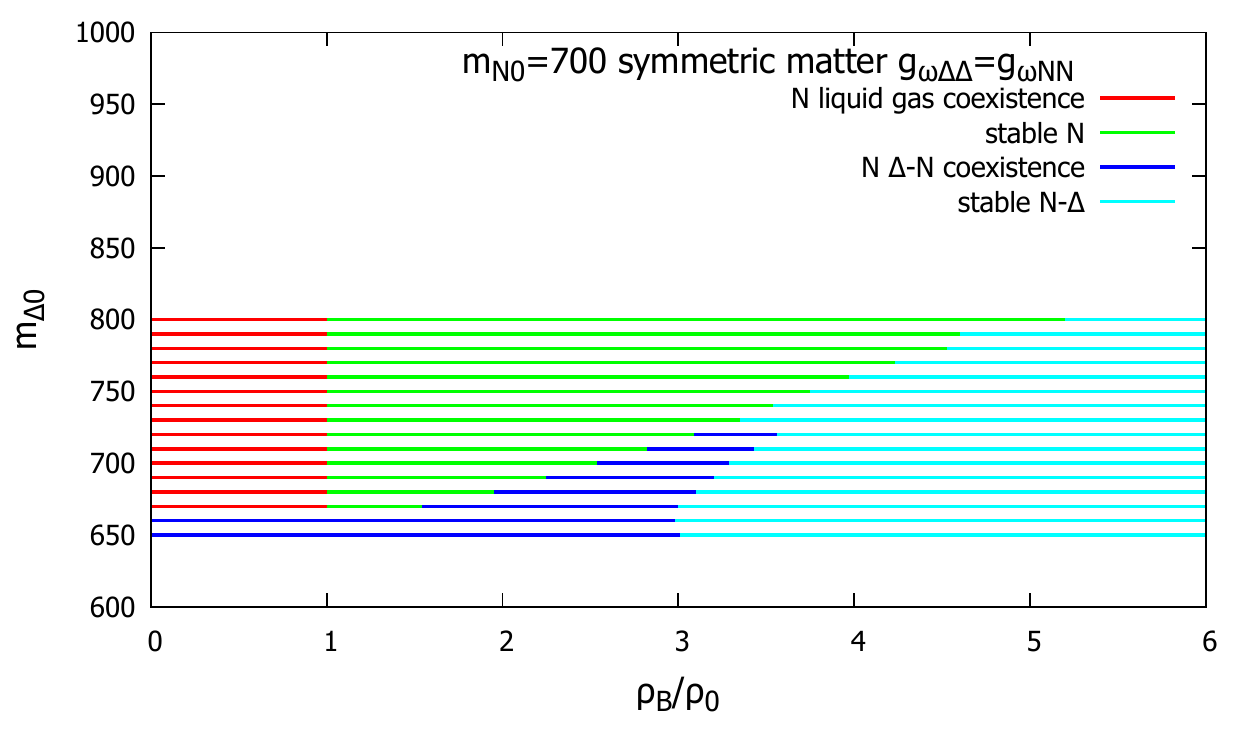}\\
 (b) \ \  \\
\vspace{-85pt}
\includegraphics[bb=0 0 480 360,width=10cm]{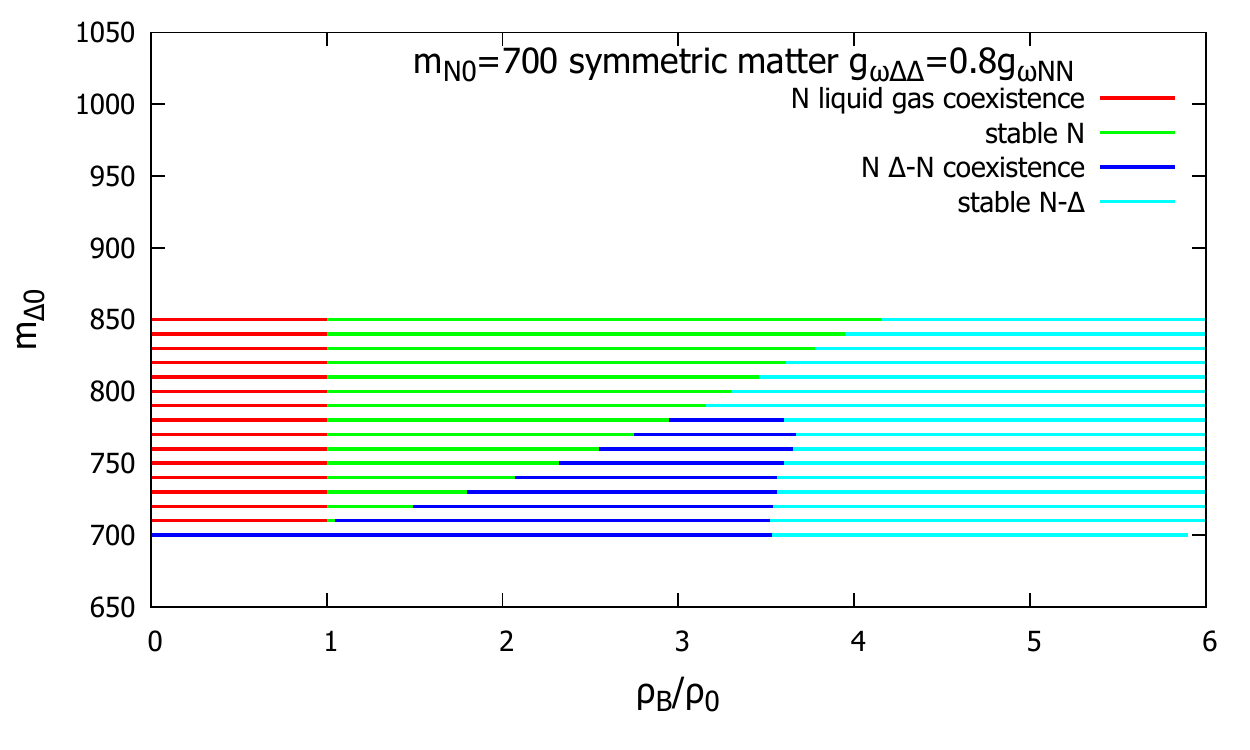}\\
 (c) \ \  \\
\vspace{-85pt}
\includegraphics[bb=0 0 480 360,width=10cm]{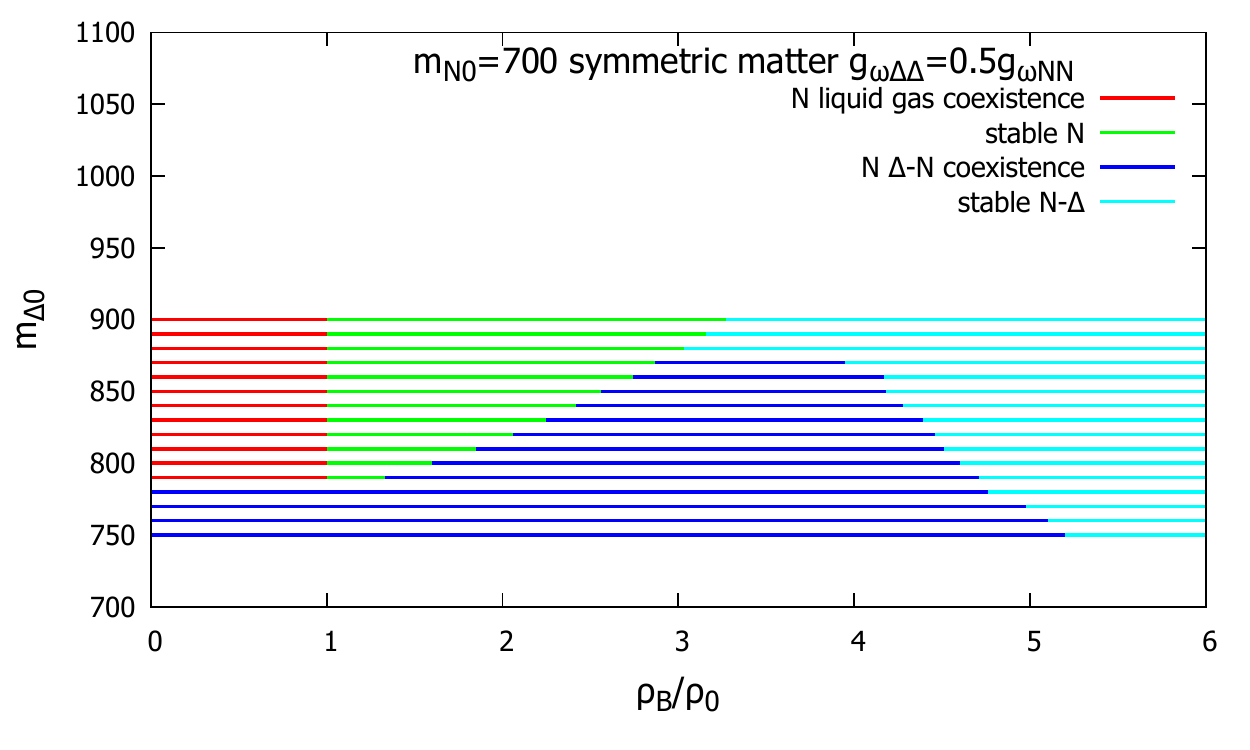}\\
\end{center}
\caption{
Matter constituents
for (a)~$g_{\omega \Delta\Delta} = g_{\omega NN}$,
(b)~$g_{\omega \Delta\Delta} = 0.8 g_{\omega NN}$  and (c)~$g_{\omega \Delta\Delta} = 0.5 g_{\omega NN}$ with the fixed value of  $m_{N 0} = 700$\,MeV
in symmetric matter.
The horizontal axis shows the baryon number density scaled  by the normal nuclear matter density, and the vertical axis shows the value of  $m_{\Delta 0}$.
The red area indicates the  ``$N$ liquid-gas coexistence'',
the green area the
  ``stable $N$ matter'',
the blue area the
 ``coexistence of $\Delta$-$N$ matter and nuclear matter'',
the cyan area the
  ``stable $N$-$\Delta$ matter''.
}
\label{phase 700}
\end{figure}

In Fig.~\ref{phase 700} we show
 matter constituents
for different choices of $g_{\omega \Delta\Delta}$ with fixed   $m_{N 0} = 700$\,MeV.
This shows that the phase structures are similar to the ones for $m_{N0}=500$\,MeV,
but the critical densities for the  change of matter constituents
depend on the choice of parameters.
Comparing Figs.~\ref{phase 500 1}-\ref{phase 500 2} with Fig.~\ref{phase 700},
we  also observe that larger $m_{N0}$ tends to lower the transition density to the
``stable $N$-$\Delta$ matter''.
In Ref.\cite{Li:1997yh}, it was shown that the onset density of $\Delta$ matter is larger for larger $m_{N}^\ast/m_N$, where $m_N^\ast$ is the effective nucleon mass at $\rho_B = \rho_0$ and $m_N$ is the mass at vacuum.
In the present analysis, even when we change the value of the chiral invariant mass $m_{N0}$, the effective nucleon mass at $\rho_B = \rho_0$ is intact.  On the other hand, the effective mass for $\rho_B > \rho_0$ is larger for larger $m_{N0}$ as shown in Fig.~\ref{fig:Nmass}.
Then, one may say that the change of the onset density against the change of $m_{N0}$ in the present analysis is consistent with the one obtained in Ref.~\cite{Li:1997yh}.
We would like to note that neither $N^\ast(1535)$ nor $\Delta(1700)$ appears in density region below $6\,\rho_0$.

\begin{figure}[htbp]
\begin{center}
 (a) \ \  \\
\vspace{-85pt}
\includegraphics[bb=0 0 480 360,width=10cm]{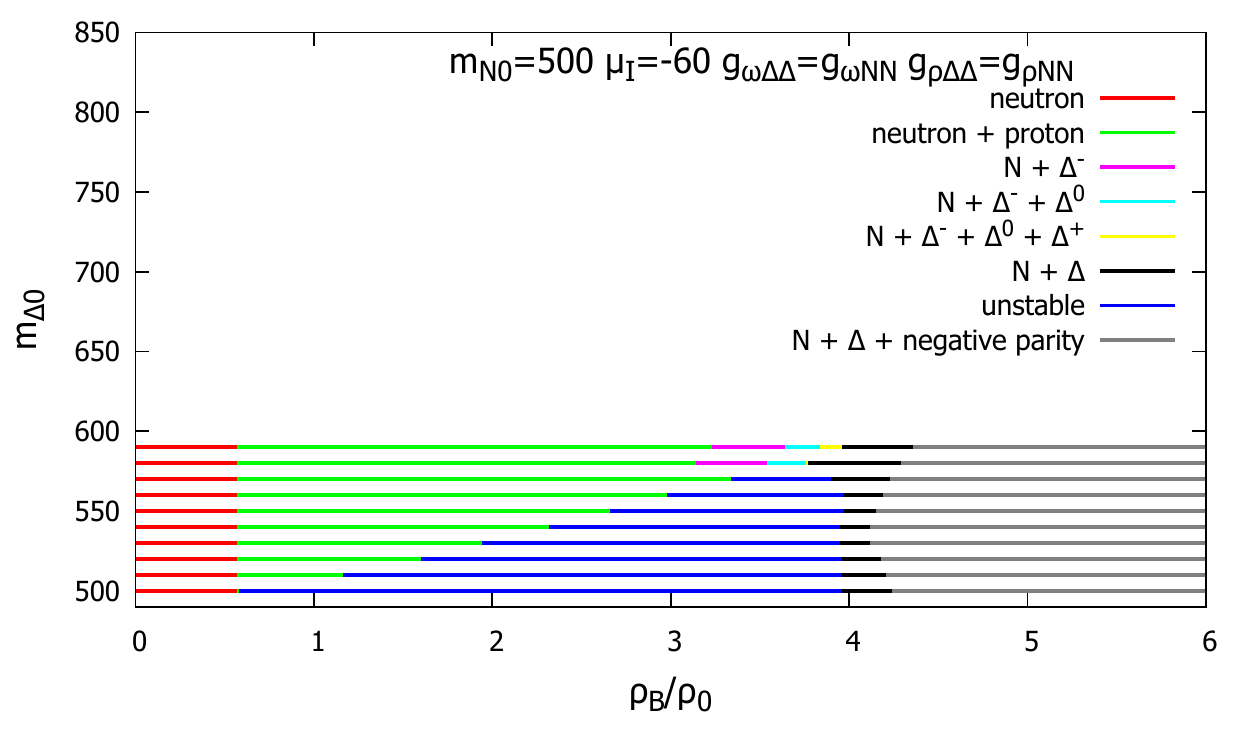}\\
 (b) \ \  \\
\vspace{-85pt}
\includegraphics[bb=0 0 480 360,width=10cm]{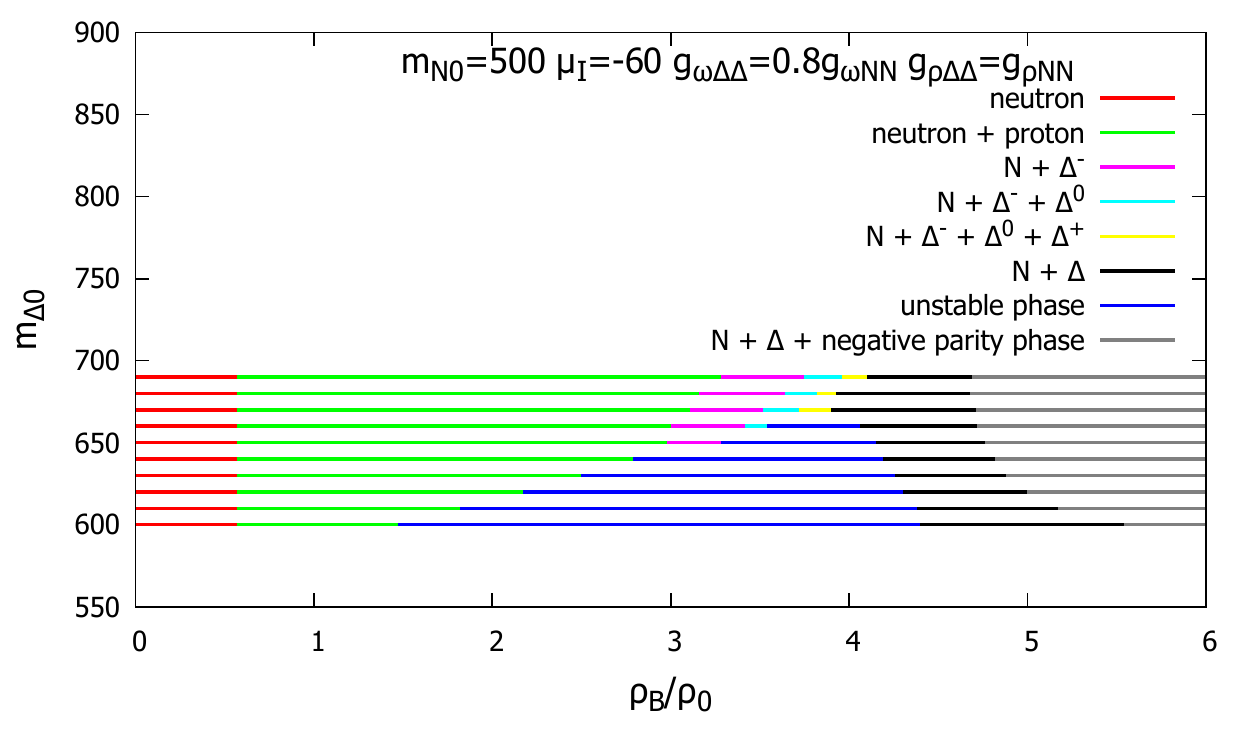}\\
 (c) \ \  \\
\vspace{-85pt}
\includegraphics[bb=0 0 480 360,width=10cm]{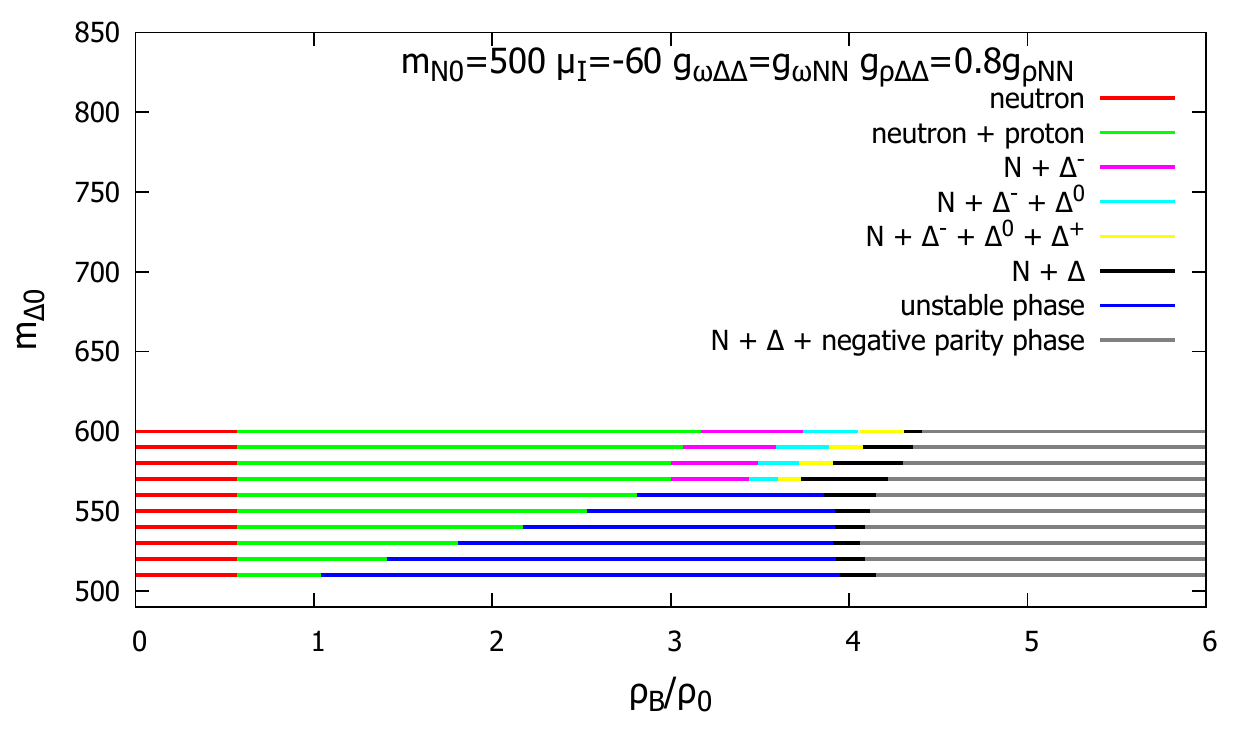}\\
\end{center}
\caption{
Matter constituents
for (a)~$(g_{\omega \Delta\Delta}/g_{\omega NN},g_{\rho\Delta\Delta}/g_{\rho NN}) = (1,1)$,
(b)~$(g_{\omega \Delta\Delta}/g_{\omega NN}, g_{\rho\Delta\Delta}/g_{\rho NN}) = (0.8,1)$
and  (c)~$(g_{\omega \Delta\Delta}/g_{\omega NN},g_{\rho\Delta\Delta}/g_{\rho NN}) = (1,0.8)$
with the fixed value of  $m_{N 0} = 500$\,MeV
in asymmetric matter with $\mu_I = - 60 $\,MeV.
The horizontal axis shows the baryon number density scaled  by the normal nuclear matter density, and the vertical axis shows the value of  $m_{\Delta 0}$.
The red area indicates the  ``neutron matter'',
the green area the
  ``stable $N$ matter'',
the blue area the  ``coexistence of $\Delta$-$N$ matter and nuclear matter'',
the black area the  ``stable $N$-$\Delta$ matter''.
In the pink area the matter is created from $\Delta^-$ and the nucleon, in the cyan area there exists the $\Delta^0$ in addition, and the $\Delta ^+$ enters the matter in the yellow area.
In the gray area, $N^\ast(1535)$ and/or $\Delta(1700)$ enter into matter.
}
\label{phase A}
\end{figure}

Let us now consider the $\Delta$ baryon in asymmetric nuclear matter.
In the following analysis, we restrict ourselves to study the property for $g_{\rho \Delta\Delta} \bar{\rho} > 0$ and $\mu_I < 0$.
From Eq.~(\ref{Delta eff mass}), the effective mass of $\Delta^-$ becomes smallest, and Eq.~(\ref{Delta chemical}) shows that the chemical potential for $\Delta^-$ is the largest.  Then, one can easily see that $\Delta^-$ enters the matter first among the four $\Delta$ baryons
 as in Refs.~\cite{Lavagno:2012bn,Schurhoff:2010ph,Drago:2014oja,Cai:2015hya,Zhu:2016mtc}.
In Fig.~\ref{phase A}, we show some examples of
 matter constituents
in asymmetric dense  matter for $\mu_I = - 60$\,MeV, by taking
$(g_{\omega \Delta\Delta}/g_{\omega NN},g_{\rho\Delta\Delta}/g_{\rho NN}) = (1,1)$,
$(g_{\omega \Delta\Delta}/g_{\omega NN},g_{\rho\Delta\Delta}/g_{\rho NN}) = (0.8,1)$
and
 $(g_{\omega \Delta\Delta}/g_{\omega NN},g_{\rho\Delta\Delta}/g_{\rho NN}) = (1,0.8)$
with fixed  $m_{N0} = 500$\,MeV
and $g_{\rho \Delta\Delta} = g_{\rho NN}$.
This shows that,
for
large values of $m_{\Delta 0}$,
the  ``stable $\Delta^-$-$N$ matte''
 indicated by the pink area appears around  $\rho_B/\rho_0 \sim 3$-$3.5$,
in which $\Delta^-$ baryon enters into the matter in addition to the  nucleon.
When the density is increased,  $\Delta^0$ and $\Delta^+$ enter as shown by the cyan and yellow areas, respectively.
The values of onset density in this smooth appearance of $\Delta$ matter are smaller than the ones in symmetric matter.
For small values of $m_{\Delta 0}$,
 on the other hand,
there exists the  ``coexistence of $\Delta$-$N$ matter and nuclear matter''
indicated by blue area.
The values of onset density are similar to the ones in symmetric matter. 
For any value of $m_{\Delta 0}$, the ``stable $N$-$\Delta$ matter''
appears in the high density region around
$\rho_B/\rho_0 \sim 4$,
where all four  $\Delta$ baryons exist in the matter.
The onset density of the ``stable $N$-$\Delta$ matter'' is smaller than in symmetric matter.
From the above, one can say that $\Delta$ matter appears in the smaller density region for asymmetric matter than for symmetric matter.
This qualitative feature seems similar to the one shown in Ref.~\cite{Lavagno:2012bn}.

We should stress that
similarly to matter constituents for symmetric matter, we observe the appearance of  the chiral partners $N^\ast(1535)$ and $\Delta(1700)$ for $\rho_B/\rho_0 \sim 4$ - $5$, reflecting the partial chiral symmetry restoration.
The values of the onset density for asymmetric matter are also smaller than those for symmetric matter.

\begin{figure}[htbp]
\begin{center}
 (a) \ \  \\
\vspace{-85pt}
\includegraphics[bb=0 0 480 360,width=10cm]{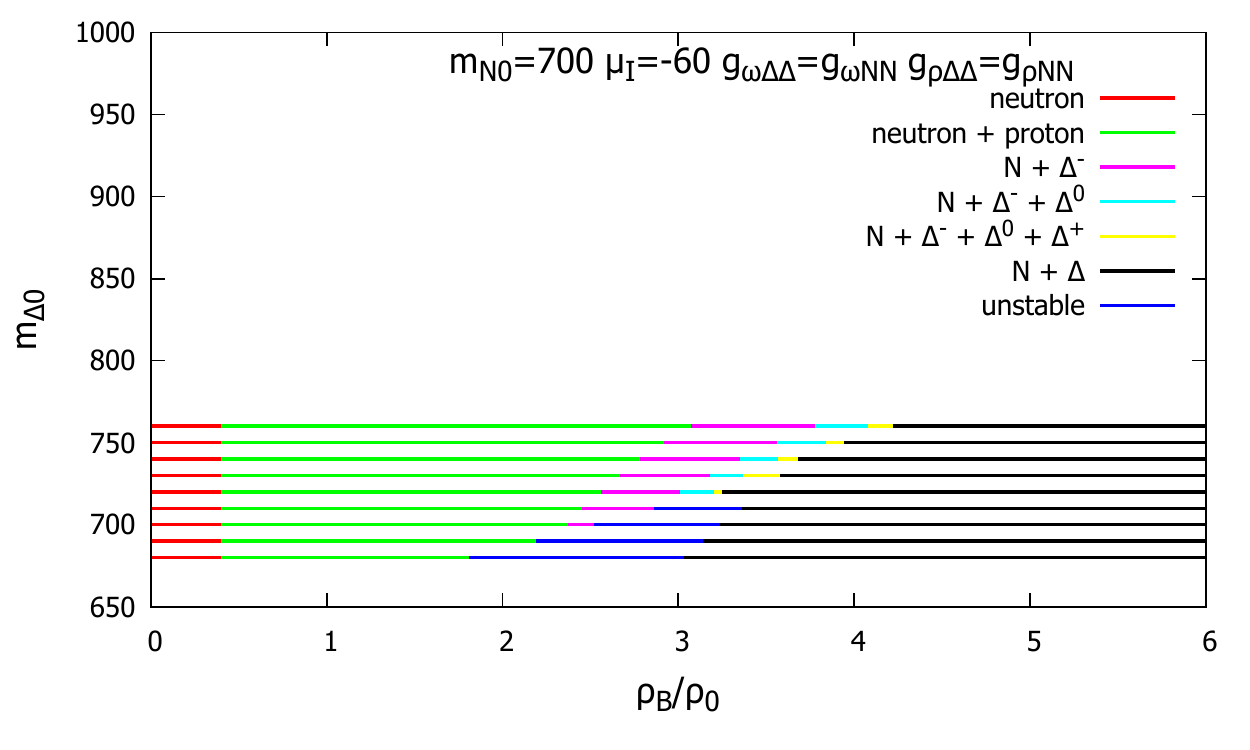}\\
 (b) \ \  \\
\vspace{-85pt}
\includegraphics[bb=0 0 480 360,width=10cm]{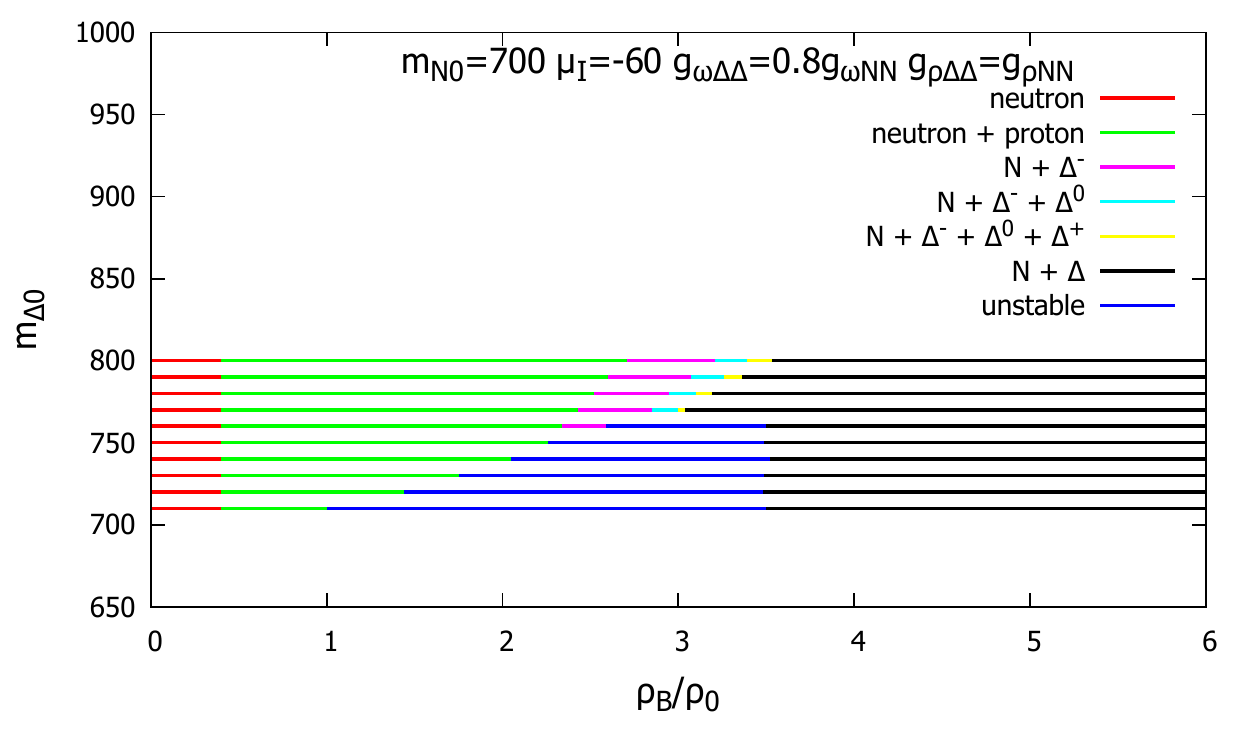}\\
 (c) \ \  \\
\vspace{-85pt}
\includegraphics[bb=0 0 480 360,width=10cm]{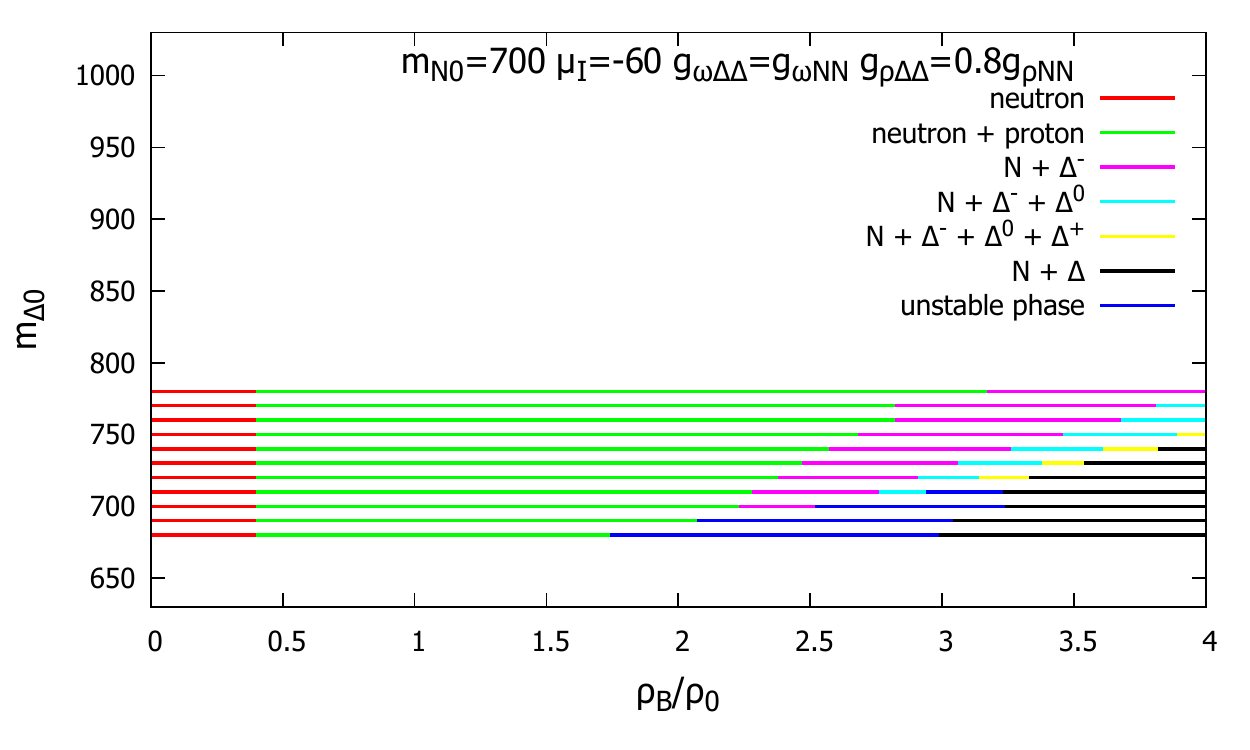}\\
\end{center}
\caption{
Matter constituents
for (a)~$(g_{\omega \Delta\Delta}/g_{\omega NN},g_{\rho\Delta\Delta}/g_{\rho NN}) = (1,1)$,
(b)~$(g_{\omega \Delta\Delta}/g_{\omega NN}, g_{\rho\Delta\Delta}/g_{\rho NN}) = (0.8,1)$
and  (c)~$(g_{\omega \Delta\Delta}/g_{\omega NN},g_{\rho\Delta\Delta}/g_{\rho NN}) = (1,0.8)$
with the fixed value of  $m_{N 0} = 700$\,MeV
in asymmetric matter with $\mu_I = - 60 $\,MeV.
The horizontal axis shows the baryon number density scaled  by the normal nuclear matter density, and the vertical axis shows the value of  $m_{\Delta 0}$.
The red area indicates the  ``neutron matter'',
the green area the
 ``stable $N$ matter'',
the blue area the  ``coexistence of $\Delta$-$N$ matter and nuclear matter'',
the black area the  ``stable $N$-$\Delta$ matter''.
In the pink area the matter is created from $\Delta^-$ and the nucleon, in the cyan area there exists the $\Delta^0$ in addition, and the $\Delta ^+$ enters the matter in the yellow area.}
\label{phase A 2}
\end{figure}

In Fig.~\ref{phase A 2}, we plot matter constituents for $m_{N0} = 700$\,MeV.

Comparing this with Fig.~\ref{phase A}, we observe that the values of the onset 
density for the appearance of $\Delta$ matter are smaller than for $m_{N0} = 500\,$MeV, similarly to symmetric matter.
From Figs.~\ref{phase 500 1}-\ref{phase 700} and Figs.~\ref{phase A}-\ref{phase A 2} we observe that the phase structures changes significantly from symmetric to asymmetric matter.

\clearpage

\section{Chiral structure}
\label{sec:chiral}

So far we have considered the
constituents of matter
of our model with the focus on $\Delta$ matter.
In this section, we study how the chiral structure is affected by the existence of $\Delta$ matter.

\begin{figure}[!tbp]
\begin{center}
\vspace{-85pt}
\includegraphics[bb=0 0 480 360,width=10cm]{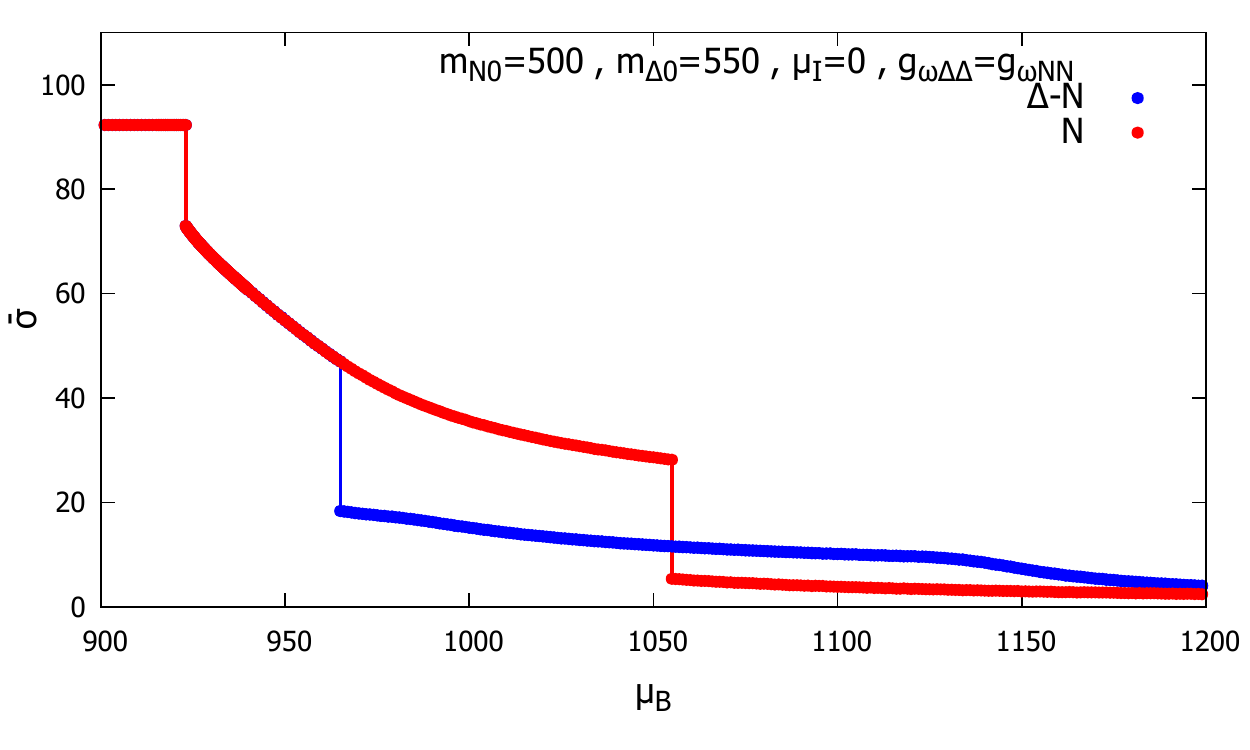}\\
\end{center}
\caption{
Chemical potential dependence of the chiral condensate $\sigma$.
The red curve shows the one with assuming no $\Delta$ in matter.
Horizontal axis shows the baryon number chemical potential in unit of MeV, while vertical axis shows the value of the chiral condensate $\sigma$  in unit of MeV.
The parameters are chosen as $m_{N0}=500$\,MeV, $m_{\Delta 0} = 550$\,MeV
and $g_{\omega \Delta\Delta} = g_{\omega NN}$.
}\label{sigma}
\end{figure}

We plot the baryon chemical potential dependence of the chiral condensate $\sigma$ in Fig.~\ref{sigma}.
Here,
the red curve is obtained without $\Delta$ baryons and the blue one is with $\Delta$.
The blue curve shows a jump at $\mu_B \sim 970$\,MeV, where $\Delta$ enters into the matter.
This shows that the partial restoration of chiral symmetry is accelerated by the existence of $\Delta$. Then, at the chemical potential slightly above $970$\,MeV, $N^\ast(1535)$ appears in matter.
 For the red curve (without $\Delta$), $N^\ast(1535)$ enters at $\mu_B \sim 1060$\,MeV, which was identified with the first order chiral phase transition in Ref.~\cite{Motohiro:2015}.
For the blue curve (with $\Delta$), on the other hand, the jump at $\mu_B \sim 970$\,MeV is not the chiral phase transition, but implies the appearance of $\Delta$ matter.
Actually, the chiral phase transition is cross-over transition which occurs at $\mu_B \sim 1140$\,MeV.
This implies that the critical chemical potential is increased when the existence of $\Delta$ is considered. In other word, the chiral symmetry restoration itself is delayed by $\Delta$ matter.

\begin{figure}[!tbp]
\begin{center}
\vspace{-85pt}
\includegraphics[bb=0 0 480 360,width=10cm]{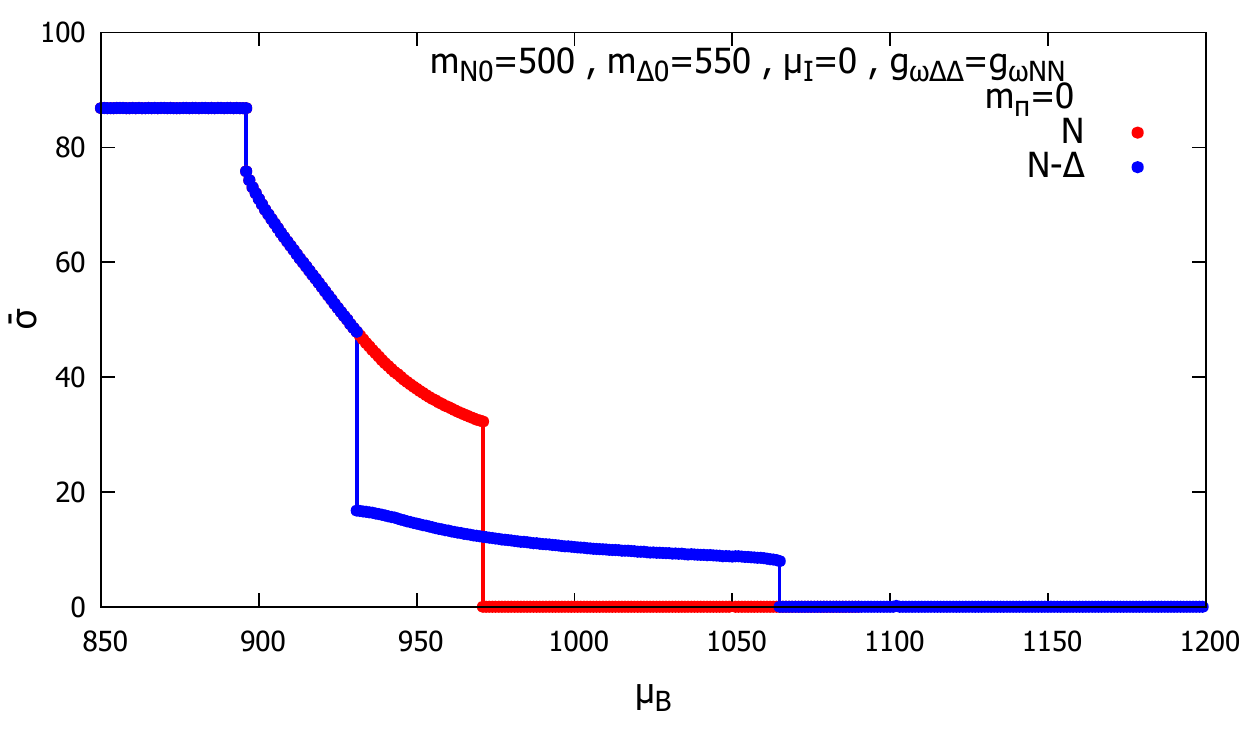}\\
\end{center}
\caption{
Chemical potential dependence of the chiral condensate $\sigma$ in the chiral limit.
Horizontal axis shows the baryon number chemical potential in unit of MeV, while vertical axis shows the value of the chiral condensate $\sigma$ in unit of MeV.
The parameters are chosen as $m_{N0}=500$\,MeV, $m_{\Delta 0} = 550$\,MeV
and $g_{\omega \Delta\Delta} = g_{\omega NN}$.
}\label{sigma chiral}
\end{figure}

In the following, to check that the small bump around $\mu_B \sim 1140$\,MeV implies the cross-over chiral phase transition, we investigate the phase structure in the chiral limit where the pion is massless.
In Fig.~\ref{sigma chiral}, we plot the chemical potential dependence of the chiral condensate $\sigma$ at the chiral limit, where we set $\bar{m} = 0$ in Eq.~(\ref{V sigma}) with all the other parameters unchanged.
Here, the red curve is without $\Delta$, and the blue curve is with $\Delta$.
The red curve jumps at $\mu_B \sim 970$\,MeV, which corresponds to the first order chiral phase transition.
The blue curve jumps at $\mu_B \sim 930$\,MeV and $\mu_B \sim 1070$. Since the $\sigma$ vanishes at $\mu_B \sim 1070$\,MeV, the first order chiral phase transition occurs at this point. The first order transition at $\mu_B \sim 930$\,MeV just indicates the appearance of $\Delta$ matter.

\begin{figure}[!tbp]
\begin{center}
\vspace{-85pt}
\includegraphics[bb=0 0 480 360,width=10cm]{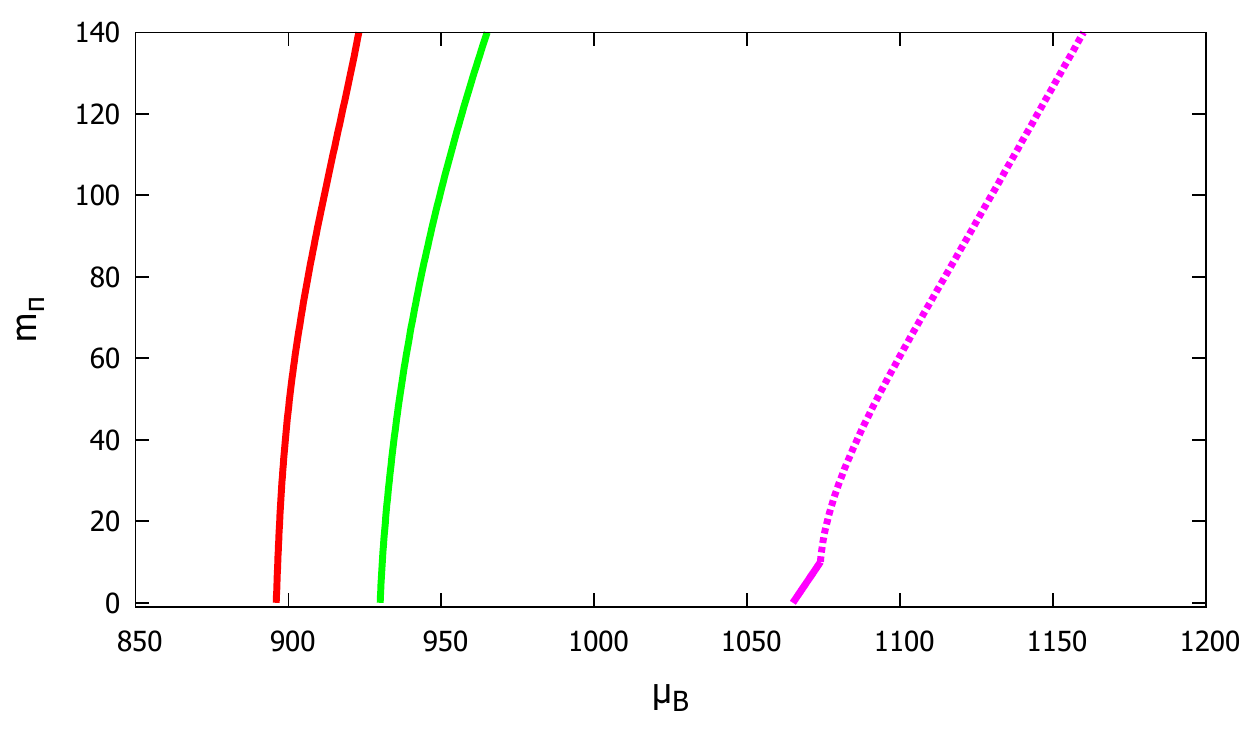}\\
\end{center}
\caption{
Chiral phase structure on $\mu_B$-$m_\pi$ plane.
Horizontal axis shows the baryon number chemical potential $\mu_B$ in unit of MeV, while vertical axis shows the value of the pion mass $m_\pi$ in unit of MeV.
The red solid curve shows the liquid-gas phase transition for the appearance of nuclear matter.
The green solid curve shows the first order phase transition for the appearance of $\Delta$ matter.
The pink solid curve shows the first order chiral phase transition and the pink dashed curve shows the cross-over chiral phase transition.
}\label{mq phase}
\end{figure}

To show that the first order chiral phase transition at $\mu_B \sim 1070$\,MeV in Fig.~\ref{sigma chiral} is connected to the crossover at $\mu_B\sim1140$\,MeV in Fig.~\ref{sigma}, we study the chiral condensate $\sigma$ with changing the $\bar{m}$ in Eq.~(\ref{V sigma}).
In Fig~\ref{mq phase}, we draw the chiral phase structure, where the horizontal axis shows the value of $\mu_B$ and the vertical axis shows the value of $m_\pi$ corresponding to $\bar{m}$.
This shows that the first order chiral phase transition (pink solid curve) at $m_\pi = 0$ and $\mu_B \sim 1070$\,MeV is actually connected to the cross-over transition at $m_\pi = 140$\,MeV and $\mu_B \sim 1140$\,MeV.
Furthermore, two critical chemical potentials for the ordinary liquid-gas phase transition and the appearance of $\Delta$ matter also increase when $m_\pi$ is increased.

\begin{figure}[!tbp]
\begin{center}
\vspace{-85pt}
\includegraphics[bb=0 0 480 360,width=10cm]{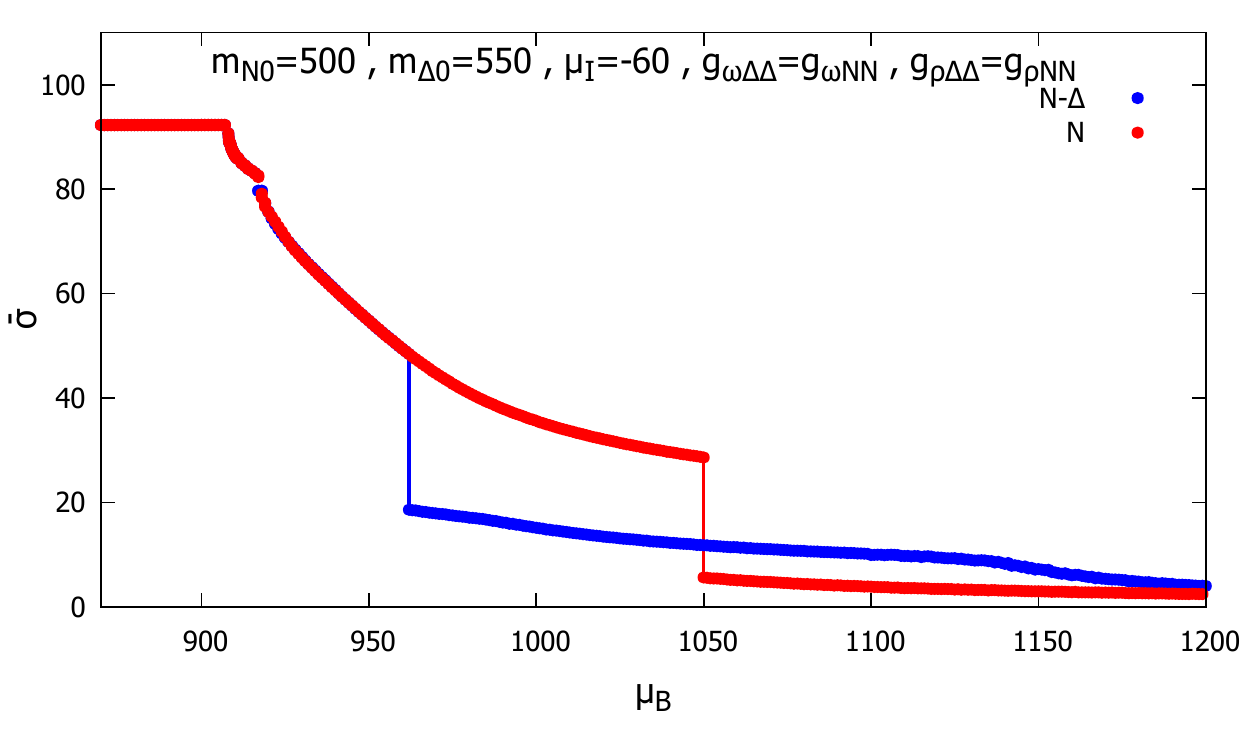}\\
\end{center}
\caption{
Chemical potential dependence of the chiral condensate $\sigma$ in asymmetric matter.
The red curve shows the one with assuming no $\Delta$ in matter.
Horizontal axis shows the baryon number chemical potential in unit of MeV, while vertical axis shows the value of the chiral condensate $\sigma$  in unit of MeV.
The parameters are chosen as $m_{N0}=500$\,MeV, $m_{\Delta 0} = 550$\,MeV
and $g_{\omega \Delta\Delta} = g_{\omega NN}$.
}\label{sigma A}
\end{figure}

We show the baryon chemical potential dependence of the chiral condensate in asymmetric matter in Fig.~\ref{sigma A}.
The qualitative structure is quite similar to the one for symmetric matter in Fig.~\ref{sigma}, except that the liquid-gas phase transition around $\mu_B \sim 970$\,MeV is changed to cross-over.

\begin{figure}[!tbp]
\begin{center}
\vspace{-85pt}
\includegraphics[bb=0 0 480 360,width=10cm]{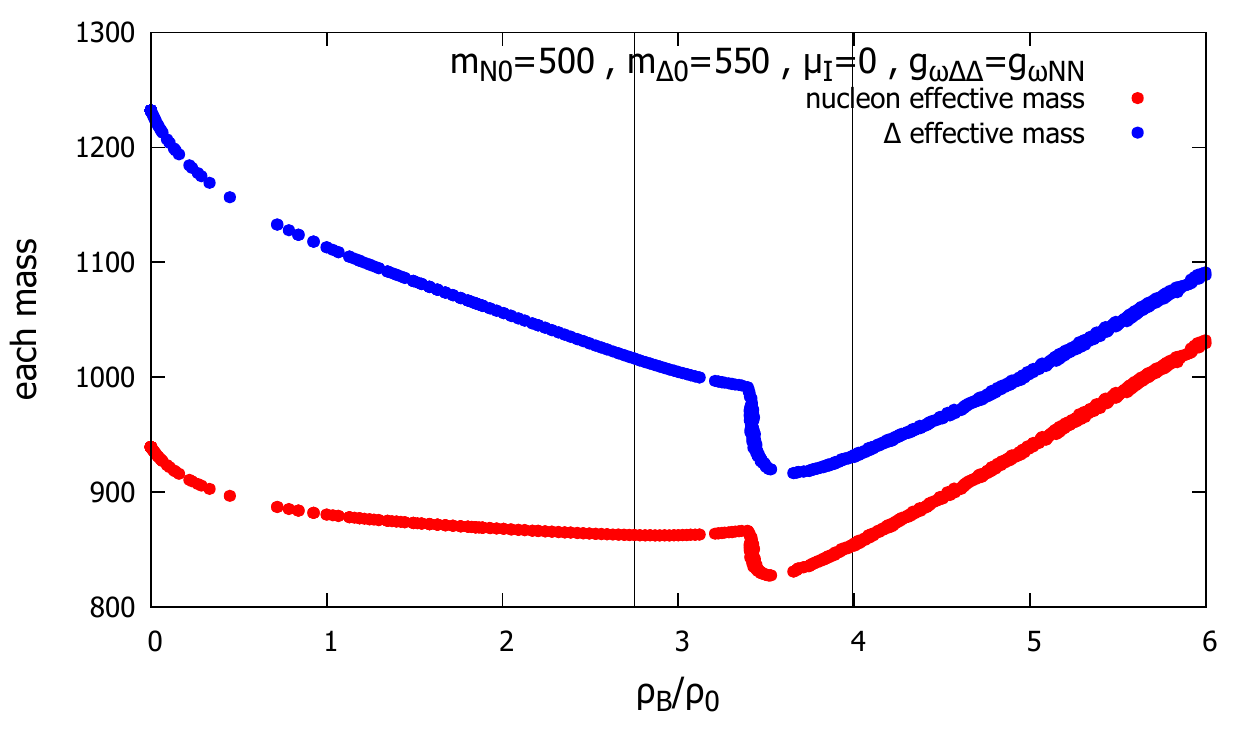}\\
\end{center}
\caption{Density dependence of the effective masses of $N(939)$ (red curve) and $\Delta(1232)$ (blue curve) in symmetric matter  for $(m_{N0},m_{\Delta0}) = (500,550)$\,MeV.
Horizontal axis shows the baryon number density $\rho_B$ normalized by the normal nuclear matter density $\rho_0$, and the vertical axis shows the value of masses in unit of MeV.  Two vertical lines at  $\rho_B/\rho_0 \sim 2.7$ and $4$
indicate that the region between them is in the  ``coexistence of $\Delta$-$N$ matter and nuclear matter''.
}\label{ef masses}
\end{figure}

At the last part of this section, we show some characteristics in other physical quantities associated with the chiral phase structure.
In Fig.~\ref{ef masses}, we plot the density dependence of  effective masses of $N$ and $\Delta$ for $(m_{N0},m_{\Delta0}) = (500,550)$\,MeV.
The region between two vertical lines at  $\rho_B/\rho_0 \sim 2.7$ and $4$
 is in the  ``coexistence of $\Delta$-$N$ matter and nuclear matter'',
 as seen in Fig.~\ref{phase 500 1}.
Figure~\ref{ef masses} shows that, in this  coexistence region,
both the effective masses of $N(939)$ and $\Delta(1232)$ suddenly change their values associated with the appearance of the $\Delta$ baryons in matter.

\begin{figure}[htbp]
\begin{center}
\vspace{-85pt}
\includegraphics[bb=0 0 480 360,width=10cm]{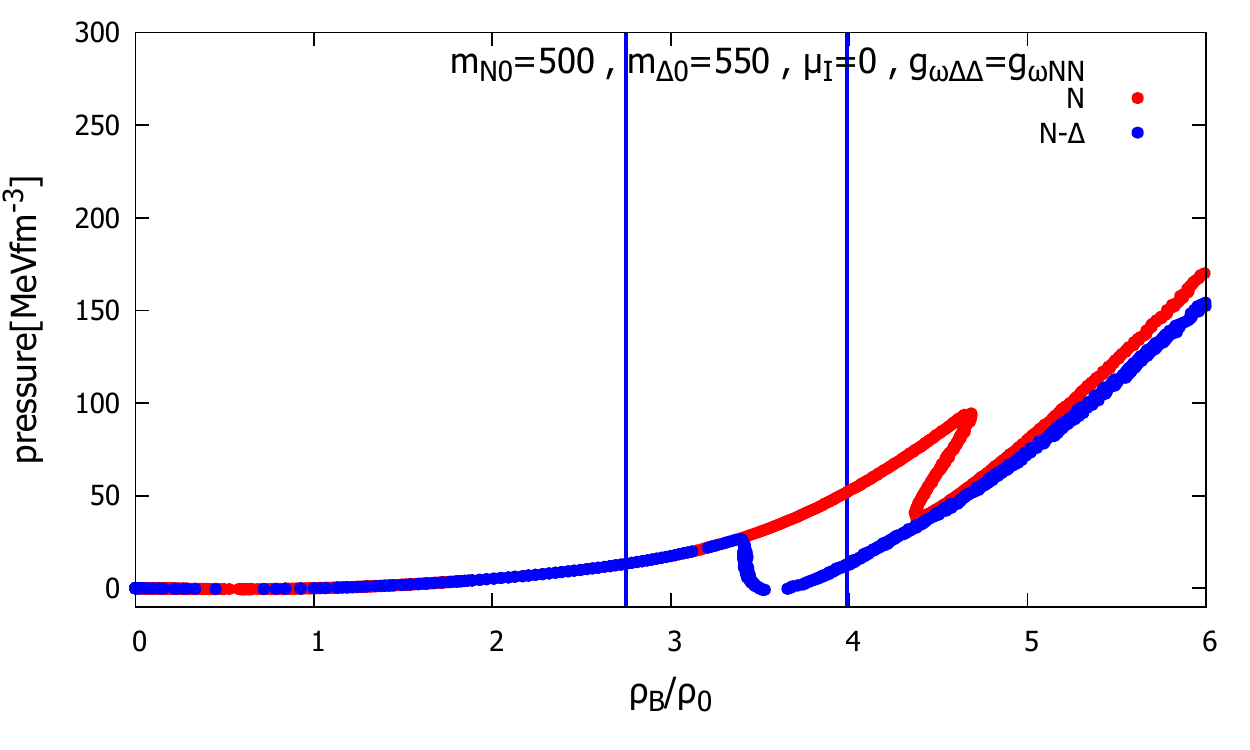}\\
\end{center}
\caption{Density dependence of pressure in symmetric matter (green curve)
for $(m_{N0},m_{\Delta0}) = (500,550)$\,MeV
The red curve shows the pressure when we assume no $\Delta$ baryons in matter.
The horizontal axis shows the baryon number density $\rho_B$ normalized by the normal nuclear matter density $\rho_0$, and the vertical axis shows the pressure in unit of MeV$\cdot$fm$^{-3}$.
The parameters are chosen as  $m_{N0}=500$\,MeV, $m_{\Delta0}=550$\,MeV
and $g_{\omega \Delta\Delta} = g_{\omega NN}$.  Two green vertical lines at
 $\rho_B/\rho_0 \sim 2.7$ and $4$
indicate that the region between them is in the
 ``coexistence of $\Delta$-$N$ matter and nuclear matter''.
For red curve, the density region of $\rho_B/\rho_0 \simeq 4.3$ - $4.7$ is unstable since $N^\ast(1535)$ enters into matter in this region.
}\label{fig:press}
\end{figure}

We plot the density dependence of the pressure in symmetric matter in Fig.~\ref{fig:press} by the blue curve.
For comparison, we show the pressure obtained by assuming that only nucleons exist in matter by the red curve.
The region between two vertical lines at  $\rho_B/\rho_0 \sim 2.7$ and $4$
 is in the  ``coexistence of $\Delta$-$N$ matter and nuclear matter."
Figure~\ref{fig:press} shows that the pressure indicated by the blue curve suddenly changes to decrease around $\rho_B/\rho_0=3.5$,
where the $\Delta$ baryons enter  the matter.
 Furthermore,
$N^\ast(1535)$ enters into matter slightly above $\rho_B/\rho_0 \sim 4$, reflecting the partial chiral symmetry restoration.
For red curve, the density region of $\rho_B/\rho_0 \simeq 4.3$ - $4.7$ is unstable since $N^\ast(1535)$ enters into matter in this region.

\begin{figure}[htbp]
\begin{center}
\vspace{-85pt}
\includegraphics[bb=0 0 480 360,width=10cm]{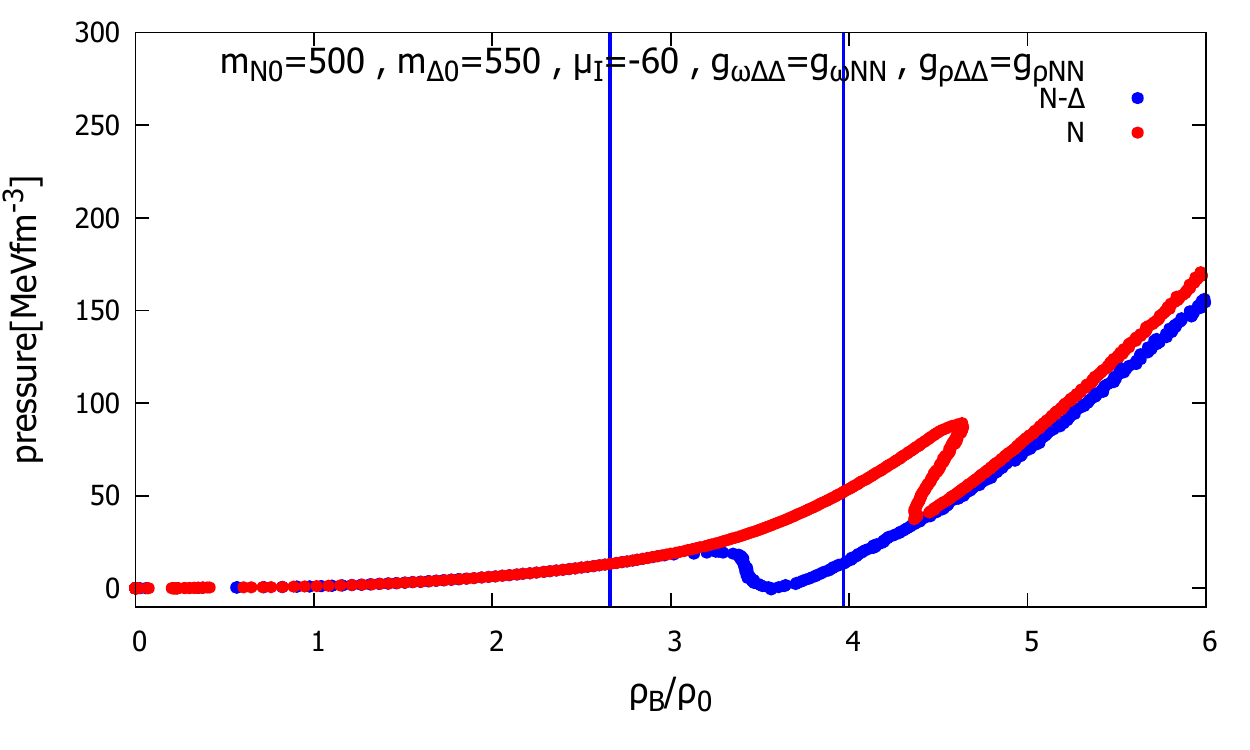}\\
\end{center}
\caption{Density dependence of pressure in asymmetric matter of $\mu_I = -60$\,MeV (blue curve).
The red curve shows the pressure when we assume no $\Delta$s baryons in matter.
The horizontal axis shows the baryon number density $\rho_B$ normalized by the normal nuclear matter density $\rho_0$, and the vertical axis shows the pressure in unit of MeV$\cdot$fm$^{-3}$.
The parameters are chosen as  $m_{N0}=500$\,MeV, $m_{\Delta0}=550$\,MeV
and $g_{\omega \Delta\Delta} = g_{\omega NN}$.  Two vertical lines at  $\rho_B/\rho_0 \sim 2.7$ and $4$
indicate that the region between them is in the  ``coexistence of $\Delta$-$N$ matter and nuclear matter''.
For red curve, the density region of $\rho_B/\rho_0 \simeq 4.3$ - $4.7$ is unstable since $N^\ast(1535)$ enters into matter in this region.
}\label{fig:pressA}
\end{figure}

We plot the density dependence of the pressure in asymmetric matter in Fig.~\ref{fig:pressA}.
The region between two vertical lines at  $\rho_B/\rho_0 \sim 2.7$ and $4$
is in the  ``coexistence of $\Delta$-$N$ matter and nuclear matter''.
In this region there is a jump around  $\rho_B/\rho_0 \sim 3.3$ where $\Delta$s
enter  the matter.
The density dependence of pressure in Figs.~\ref{fig:press} and \ref{fig:pressA} show that the pressure of $N$-$\Delta$ matter is smaller than that of the ordinary $N$ matter, and that the $\Delta$ softens the equation of state as expected.

\begin{figure}[htbp]
\begin{center}
\vspace{-85pt}
\includegraphics[bb=0 0 480 360,width=10cm]{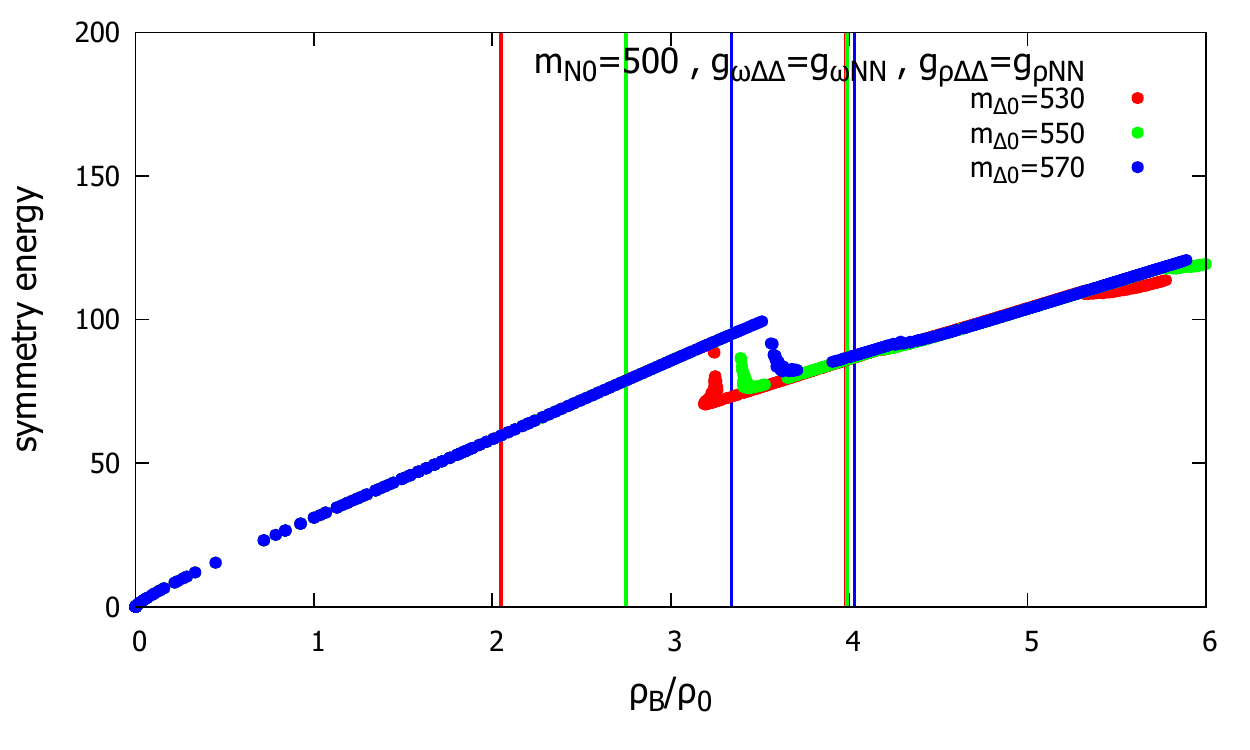}\\
\end{center}
\caption{Density dependence of symmetry energy.
Two red vertical lines at  $\rho_B/\rho_0 \sim 2$ and $4$
indicate that the region between them is in the  ``coexistence of $\Delta$-$N$ matter and nuclear matter'' for $m_{\Delta0}=530$\,MeV.
Two green vertical lines at  $\rho_B/\rho_0 \sim 2.7$ and $4$ and two blue lines at $\rho_B/\rho_0 \sim 3.3$ and $4$
indicate the coexistence  region for $m_{\Delta0} = 550$\,MeV and $m_{\Delta0}=570$\,MeV, respectively.
}\label{fig:symE}
\end{figure}

In Fig.~\ref{fig:symE}, we plot the density dependence of the symmetry energy.
With fixing  $m_{N0}=500$\,MeV,
$g_{\omega \Delta\Delta} = g_{\omega NN}$ and $g_{\rho \Delta\Delta} = g_{\rho NN}$, we use three inputs for the chiral invariant mass of the $\Delta$:  $m_{\Delta0} = 530$\,MeV (red curve), $550$\,MeV (green curve) and $570$\,MeV (blue curve).
The region between two red vertical lines at  $\rho_B/\rho_0 \sim 2$ and $4$
is in the  ``coexistence of $\Delta$-$N$ matter and nuclear matter'' for $m_{\Delta0}=530$\,MeV,
while the region between two green vertical lines at  $\rho_B/\rho_0 \sim 2.7$ and $4$  is for $m_{\Delta0}=550$\,MeV, and blue lines at $\rho_B/\rho_0 \sim 3.3$ and $4$ is for $m_{\Delta0} = 570$\,MeV.
In the coexistence region, there are a few solutions for one value of density, then there are a few values of symmetry energy for some density in Fig.~\ref{fig:symE}.
Figure~\ref{fig:symE} shows that the symmetry energy suddenly changes its value around the density where  $\Delta$ enters the matter.


\section{summary and discussion}\label{summary}

In the framework of an extended parity doublet model with $\Delta$ baryons, we studied the transition from nuclear matter to $\Delta$ matter
in terrestrial dense matter that can be created in heavy ion collisions (HIC)
 with a few hundreds A MeV.
We also investigated the effects of  the chiral invariant mass  and the isospin asymmetry on the transition to  $\Delta$ matter and on some physical quantities such as
the nuclear symmetry energy.

We explored matter constituents of cold dense matter and showed
that, in symmetric matter, $\Delta$ enters into matter at $\rho_B \sim 1$ - $4 \rho_0$, and that stable $\Delta$ matter exists for $\rho_B \gtrsim 4 \rho_0$.
We observed in symmetric dense matter that larger $m_{N0}$ tends to lower the transition density to the ``stable $N$-$\Delta$ matter''.
We also showed that the matter constituents change significantly with the finite isospin chemical potentials.
In asymmetric matter, $\Delta^-$ appears first similarly to the ones  obtained in the neutron star analyses~\cite{Schurhoff:2010ph,Drago:2014oja,Cai:2015hya,Zhu:2016mtc}.

We next investigated the chiral structure by studying the chemical potential dependence of the chiral condensate $\sigma$.
We showed that the chiral condensate rapidly decreases when $\Delta$ enters into matter, implying that the partial chiral symmetry restoration is accelerated by $\Delta$ matter.
When the chiral invariant mass of nucleon is $500$\,MeV, as a result of this acceleration,
$N^\ast(1535)$, which is the chiral partner to $N(939)$, enters into matter at the density slightly above the onset density of  stable $\Delta$ matter, signaling the partial chiral symmetry restoration.
We also showed that  the chiral symmetry restoration itself is delayed by $\Delta$ matter.

We finally calculated the effective masses, pressure and symmetry energy.
The density-dependences
of effective masses and pressure change drastically around the onset density of $\Delta$ matter, which are similar to the ones shown in Refs.~\cite{Xiang:2003qz,Schurhoff:2010ph,Zhu:2016mtc}.
We also observed a sudden change of symmetry energy around the onset density.

In the present study we do not consider any possibility of having hyperon matter because strangeness is conserved with strong interactions and
terrestrial dense matter from  heavy ion collisions at a few hundreds A MeV  sustains much short compared to the time scale of weak interactions.

It will be interesting to see how the observations made in  this study such as the transition to $\Delta$ matter affect the observables  in
heavy ion collisions such as neutron-proton collective flows and $\pi^+/\pi^-$ ratio at low and/or intermediate energy in a transport model simulation, which is relegated to our future study. Also, the fate of pion condensation with $\Delta$ matter  in heavy ion collisions will be investigated.
For the earlier discussion on the role of the $\Delta$ baryons in pion condensation, we refer to~\cite{Tripathi:1980jy,Dickhoff:1981jq}.

\subsection*{Acknowledgments}

This work was supported partly by the Rare Isotope Science Project of Institute for Basic Science funded by Ministry of Science, ICT and Future Planning and NRF of Korea (2013M7A1A1075764),
and the JSPS Grant-in-Aid for Scientific Research (C) No.~16K05345.


\end{document}